\documentclass[12pt, a4paper]{article}

\usepackage[a4paper,
            bindingoffset=0.2in,
            left=0.75in,
            right=0.75in,
            top=1in,
            bottom=1in,
            footskip=.25in]{geometry}

\usepackage{graphicx, caption}
\usepackage{xcolor}

\usepackage[utf8x]{inputenc} 
\usepackage{amsmath, amsthm} 
\usepackage[bottom]{footmisc}
\usepackage{cite}
\usepackage{graphicx}
\usepackage{longtable}
\usepackage{float}
\usepackage{lscape}
\usepackage{slashbox}
\usepackage{hyperref}
\usepackage{braket}
\usepackage{multirow}
\usepackage{subfigure}
\usepackage{algorithm}
\usepackage{algorithmic}
\usepackage{array}
\usepackage{amssymb,bm,cite,graphicx, fixmath, texdraw}
\usepackage{epsfig}
\usepackage{epstopdf}
\usepackage{subfigure} 
\usepackage{amsmath}
\usepackage{notoccite}

\usepackage{authblk}

\begin{document}

\title{Analysis of Pantheon+ supernova data suggests evidence of sign-changing pressure of the cosmological fluid }

\author[1]{A. Kaz\i m \c Caml\i bel \footnote{camlibel@tau.edu.tr}}
\author[2]{M. Akif Feyizo\u{g}lu \footnote{akif.feyizoglu@std.bogazici.edu.tr, akif.feyizoglu@gmail.com}}
\author[2]{\. Ibrahim Semiz \footnote{ibrahim.semiz@bogazici.edu.tr, ibrahim.semiz.fizik@icloud.com}}
\affil[1]{Electrical and Electronics Engineering Department, Turkish-German University, 34820 Beykoz, \. Istanbul, Turkey}
\affil[2]{Physics Department, Bo\u{g}azi\c  ci University, 34342 Bebek, \. Istanbul, Turkey}

\maketitle

\abstract{

In this work, we revisit/reinterpret/extend the model-independent analysis method (which we now call spread - luminosity distance fitting, spread-LDF) from our previous work. We apply it to the updated supernova type Ia catalogue, Pantheon+ and recent GRB compilations. The procedure allows us, using only FLRW assumption, to construct good approximations for expansion history of the universe, re-confirming its acceleration to be a robust feature. When we also assume General Relativity (“GR”), we can demonstrate, without any matter/energy model in mind, the need for (possibly nonconstant) dark energy (“GDE”). We find hints for positive pressure of GDE at $z > 1$ with implications on either the complexity of dark energy, or the validity of one of the cosmological principle, interpretation of SN Ia data, or GR.

}

\section{Introduction}

It has been more than 25 years since the discovery that the universe --with the assumptions that it is homogeneous and isotropic-- undergoes an accelerated expansion \cite{Riess, Perlmutter}. Researchers tried to explain the cosmic acceleration, either by exotic matter-energy components, collectively coined as dark energy  \cite{Li}, or by modifying the existing theories of gravity \cite{Shankaranarayanan}. In addition to those, alternative approaches, which relax the cosmological principle, tried to explain the SN Ia observations with inhomogeneities in the matter content of the universe \cite{BuchertRasanen}.

Original results from High-z Supernova Team  \cite{Riess}and Supernova Cosmology Project \cite{Perlmutter}, back in 1998 and 1999, were using only 16 and 42 high redshift SNe Ia respectively, up to  $z\sim 0.8$. The basic intuition from that point on was that the increasing number of observed SNe Ia at higher redshifts would help to distinguish between the benchmark ${\rm \Lambda}$CDM model and its alternatives. More SNe Ia observations throughout the past decades came from various surveys, using both earth-based and space telescopes, with different light curve fitting strategies, but presenting catalogues with overlapping sources. To combine those catalogues, greater compilation sets were created, which give redshifts, apparent magnitudes and/or distance moduli of the SNe Ia in a unified manner so that independent research groups can analyze them for their own cosmological models. One of the earliest compilations was the Union set \cite{Kowalski} back in 2008, which combined 307 SNe Ia up to redshifts $z\sim1.5$. That was improved in 2009, where the resulting Constitution set \cite{Hicken} presented about 400 SNe Ia in a similar range. That was followed by Union 2 \cite{Amanullah} and Union 2.1 \cite{Suzuki} updates, which involved 557 and 580 SNe Ia respectively, within similar redshift ranges as the Union set, but an increased number of SNe above $z\sim1$. In 2019, the Pantheon set was published \cite{Pantheon}. This one was a big update, mostly from the Pan-STARSS1 survey, with 1048 SNe Ia with a redshift range up to $z\sim2.3$. The most recent catalogues are Union 3 \cite{Rubin} and Pantheon+ \cite{Pantheon+} compilations, with 2087 and 1701 SNe Ia respectively. We would also like to mention the Dark Energy Survey (DES) catalogue \cite{Vincenzi}. Despite the fact that it is not a compilation from different surveys, it presents 1635 SNe Ia in the redshift range of 0.10 - 1.13.

Vast improvement in SN Ia observations over the years did not end up in favoring of one cosmological model over the others.  ${\rm \Lambda}$CDM is still strongly favorable, especially considering SN Ia and CMB observations; yet inclusion of latest BAO measurements hint a preference for evolving dark energy models \cite{Adame}. However, most of the other alternative cosmic scenarios are still relevant, in contrast to the expectations that the improvement in SN Ia observations would solve the DE problem. Explanation to this may be given as follows: First of all, new SNe at higher redshifts exhibit greater uncertainties in magnitude measurements due to technical limitations in measuring dimmer objects, which decreases the distinguishing power for those sources. Secondly, dark energy models or modified gravity theories usually have enough number of free parameters, which allows them to fit the SNe magnitudes within acceptable margins. At that point, finer statistical analysis of data and employing model-independent analysis methods became more important for getting a more accurate understanding of the expansion characteristics of the universe.

Model-independent expansion diagnostics \cite{Sahni, Sahni2},  and cosmographic analysis methods \cite{Visser} became important in determining the expansion characteristics of the universe \cite{Myrzakulov} and they served as alternative tools in differentiating between different dark energy models \cite{Errehymy}. More recently, calculation of those cosmographic parameters went beyond the SN Ia data with the cosmological treatment of fast radio bursts (FRB) \cite{Gao}, gamma ray bursts (GRB) \cite{Bargiacchi} and Quasars \cite{BargiacchiQ}. In addition to that, researchers started using cosmography as a test of anisotropy of the universe \cite{Heinesen, Bengaly}. 

In our first paper on the subject \cite{Semiz}, where we analyzed Union 2.1 set, our main goal was to get the most information about the expansion with the least number of assumptions possible. For this, we introduced a method, which we now would like to call spread-Luminosity Distance Fitting (spread-LDF). That work showed that the analyzed SN catalogue was not  powerful enough to unequivocally tell that we were even in an accelerated expansion phase. Only with the inclusion of farther objects, such as GRB's, it was possible to tame our luminosity distance functions and estimate a transition redshift, where universe enters an accelerated phase.

We updated our analysis in a follow-up paper \cite{Akif}, where the same analysis methods were applied to Pantheon set. This work showed a significant improvement over the last one, which exhibited a consistent transition from deceleration to acceleration for most of the fitting functions, even without the inclusion of GRB's.

This work extends our analysis to one of the most recent compilations, Pantheon+ set. In the next section we present the data to be used and review the spread-LDF procedure. Section 3 presents the results of the procedure, i.e. conclusions about the expansion history of the universe in form of $\dot{a}(z)$, $\ddot{a}(z)$, and implications thereof. In section 4, we adopt General Relativity (GR), the simplest theory of gravitation, and show the necessity of Dark Energy without assuming it. However, we find the intriguing result that the pressure of the cosmic fluid changes sign around $z\approx1$. The last section summarizes the results, and elaborates on future prospects, in particular that of the sign change mentioned.

\section{The Data and the Spread-LDF Approach}

In this section, we discuss the cosmological supernova and GRB data that we will be using; and review the method, introduced in previous work and to be used in the following two sections, of constructing the time-derivatives of the scale factor from these data, assuming the cosmological principle.

\subsection{Data Sets}

\begin{figure}[h!]
\centering
\includegraphics[width=1 \columnwidth]{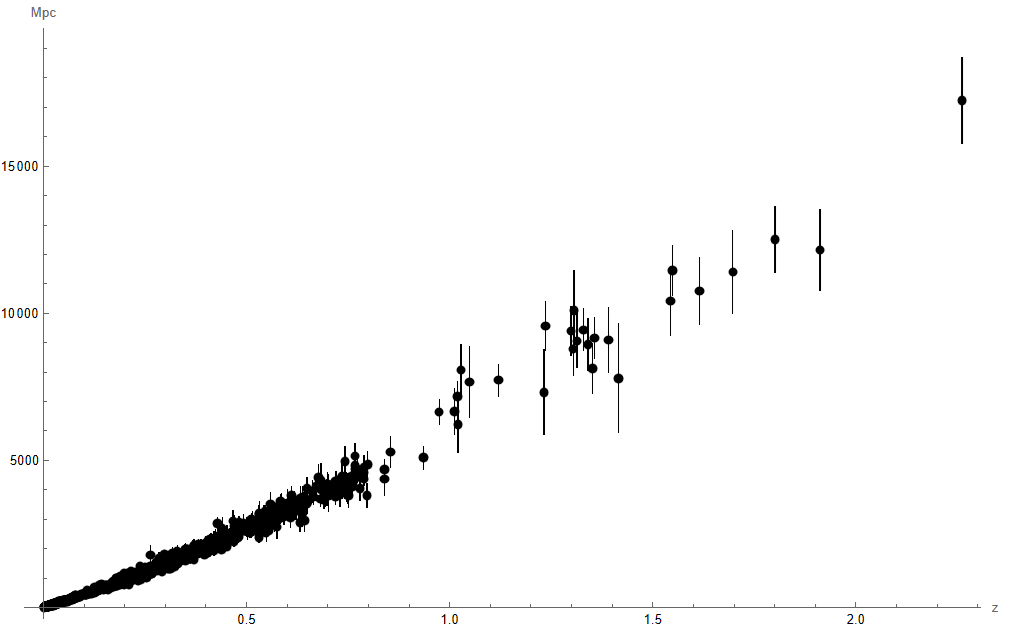}
\caption{Luminosity distance $d_L(z)$ versus redshift for Type Ia supernovae in the Pantheon+ compilation.} 
\label{Data_sn}  
\end{figure}

In this work, we primarily utilize the Pantheon+ compilation of Type Ia supernovae \cite{Pantheon+}, which provides luminosity distance measurements for over 1500 supernovae across the redshift range $0.001<z<2.26$. Compared to earlier datasets, Pantheon+ offers substantially reduced statistical and systematic uncertainties, allowing for a more precise reconstruction of the expansion history. A plot of the Pantheon+ data is shown in Figure \ref{Data_sn}. When compared to Pantheon data used in previous work (\cite{Akif}, Figure 1 in same format), wider $z$-range and smaller errors are apparent.
To further stabilize the high-redshift behavior of the reconstructed functions, we supplement the supernova sample with a set of Gamma-Ray Burst (GRB) \cite{GRB} luminosity distance measurements extending up to $z\sim 8$. Although GRB distance indicators are not completely model-independent due to calibration relations, the GRB dataset constitutes only approximately 10 percent of the total number of data points. Their inclusion primarily serves to anchor the high-redshift fits without significantly affecting the results in the low-redshift regime where the supernova data dominate. Additionally this procedure suppresses numerical oscillations in second derivatives.

However we noticed that the inclusion of GRB’s, despite their low numbers and weights (large errors) produced a noticeable change in the transition redshift, unlike in our previous work  \cite{Semiz, Akif}. Closer inspection revealed that the luminosity distance-redshift initial slope of the new dataset \cite{GRB} was about 1.7 times smaller than that of the old one \cite{GRB2},  another aspect of the model-dependence of the GRB-data. Hence we decided to scale the data to make the slope match the average of the slopes of the two datasets. It turned out that with this correction, inclusion of GRB data did not affect the transition redshift appreciably.

\begin{figure}[h!]
\centering
\includegraphics[width=1 \columnwidth]{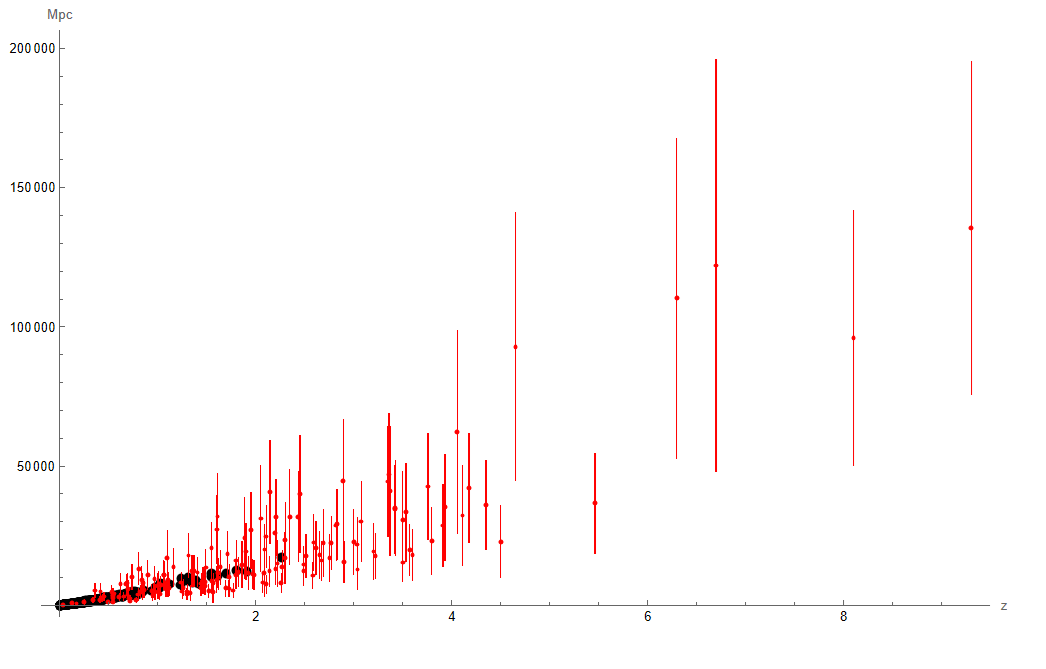}
\caption{Luminosity distance $d_L(z)$ as a function of redshift for the full observational dataset. Black points represent Type Ia supernovae from the Pantheon+ compilation; red points represent the rescaled GRB dataset. The GRBs extend the redshift range beyond the supernova domain.} 
\label{data} 
\end{figure}

The Pantheon+ collaboration provides standardized distance moduli $\mu(z)$ for each supernova, corrected using the Tripp \cite{Tripp} method. The zero-point calibration, including the fiducial absolute magnitude $M$, is based on Cepheid-host distances from the SH0ES 2021 analysis \cite{shoes}. Although the absolute scale of the luminosity distance depends on this calibration, the overall shape of the $d_L(z)$ curve, which governs the reconstruction of the expansion history, is largely unaffected. We adopt the provided $\mu(z)$ values directly and convert them to luminosity distances using the standard relation
\begin{equation}
d_L(z)=10^{\mu(z)/5-5} \textnormal{Mpc}.
\end{equation}
A combined plot of the Pantheon+ and rescaled GRB data is given in Figure \ref{data}, illustrating the wider redshift coverage of the GRB data and the relation of the two observational datasets used in our analysis.

\subsection{The Idea: From Observations Back to Expansion and Acceleration}

The idea introduced \cite{Semiz} and reapplied \cite{Akif} in previous work is that the observations (Figs. \ref{Data_sn} \& \ref{data}) give us in principle the function $d_L(z)$ and the procedure of going from the function $a(t)$, presumably found starting from a matter/energy model and a theory of gravity, to the function $d_L(z)$ can in principle be inverted; enabling in principle the determination of the scale factor’s history, the $a(t)$ function, without assumptions on matter/energy or theory of gravity.

A critical step is the relation
   
\begin{equation}\label{dtoverdz}
\frac{dt}{dz}=-\frac{1}{c(1+z)}\frac{1}{\sqrt{1-\kappa \frac{d_L^2(z)}{(1+z)^2}}}\frac{d}{dz}\left(\frac{d_L(z)}{1+z}\right)
\end{equation}
shown in previous work \cite{Semiz}, which upon integration would give $t(z)$ for a determined (fitted) $d_L(z)$ function. The inversion of $t(z)$  would give $z(t)$, finally giving $a(t)$ by the well-known relation

\begin{equation}
a(t)=\frac{a_0}{1+z}.
\end{equation}
But for except the simplest possibilities for the $d_L(z)$ function, the mentioned integration and/or inversion steps are not analytically possible, so that $a(t)$ can only be found numerically. However, it $is$ possible to find $\dot{a}$ and $\ddot{a}$ $as~functions~of~z$ by application of chain rule, since $dz/dt$ is known from Eq. \ref{dtoverdz}:

\begin{equation}\label{adot}
\dot{a}(z)=\frac{da}{dt}=\frac{da}{dz}\left(\frac{dt}{dz}\right)^{-1}
\end{equation}
and

\begin{equation}\label{addot}
\ddot{a}(z)=\left(\frac{dt}{dz}\right)^{-1}\left(\frac{2 a_0}{(1+z)^3}\left(\frac{dt}{dz}\right)^{-1}+\frac{\partial}{\partial z}\left(\frac{dt}{dz}\right)\frac{a_0}{(1+z)^2}\left(\frac{dt}{dz}\right)^{-2}\right).
\end{equation}
Naturally, the Hubble parameter can also be found in terms of $z$:

\begin{equation}\label{hubble}
H(z)=\frac{\dot{a}(z)}{a(z)}=-\frac{1}{1+z}\left(\frac{dt}{dz}\right)^{-1}.
\end{equation}

In calculations \ref{adot} and \ref{addot}, we will consider 11 equally spaced values of $\kappa$ (needed in Eq. \ref{dtoverdz}) in the range $-\kappa_0 \leq \kappa \leq +\kappa_0$, where $\kappa_0 = (10,000{\rm Mpc})^{-2}$; that is, we will consider a range of spatial curvatures. Hence, when we plot functions $\dot{a}(z)$ , $\ddot{a}(z)$  or $H(z)$, we will show 11 curves for a given $d_L(z)$, color-coded by curvature.

For a reconstructed function  $A(z)$, the uncertainties can be propagated from the fitted parameters by standard linear error propagation. Specifically, the variance of $A(z)$, depending on fit parameters $\{a_i\}$ is computed as

\begin{equation}
\sigma^2_A(z)=\Sigma_i\left(\frac{\partial A}{\partial a_i}\right)^2+2 \Sigma_{i<j}\left(\frac{\partial A}{\partial a_i}\right)\left(\frac{\partial A}{\partial a_j}\right)\rm{Cov}(a_i,a_j),
\end{equation}
where $\sigma^2_{a_i}$  and Cov$(a_i,a_j)$ are the variances and covariances of the fitted parameters provided by the fitting procedure. We can then construct $1$-$\sigma$ confidence bands around the reconstructed function by plotting $A(z)\pm \sigma_A(z)$.

We believe the idea of fitting a function to the observed data, the $d_L - z$ pairs, and derive $\dot{a}(z)$, $\ddot{a}(z)$, and hence some other cosmological functions of interest analytically from the original $d_L(z)$ fit, is different enough to merit its own name. We suggest “luminosity-distance-fitting (LDF) procedure”. 

\subsection{Spreading the LDF: Towards Model-independence}

Any $\it{ansatz}$ for the fit to represent the data of Figs. \ref{Data_sn} \& \ref{data} as a function $d_L(z)$ would be equivalent to some model, which in the beginning of the previous subsection we stated we wanted to avoid. Therefore we use a range of template functions  trying to cover as much as possible of the space of functions that can reasonably be fitted. We also use a range of independent variables, since it is not a priori obvious that redshift is the best one. Our work becomes model-independent by averaging over (at least part of) the space of possible models, and independent variables.

As discussed in previous work \cite{Semiz, Akif} we use six different redshift reparameterizations, defined by

\begin{equation}
y_0=z, \quad y_1=\frac{z}{z+1}, \quad y_2=\textnormal{arctan} \frac{z}{z+1},
\end{equation}
\begin{equation}
y_4=\textnormal{arctan}\; z, \quad y_5=\textnormal{ln}(1+z), \quad y_6=u=z+1,
\end{equation}
where $y_1-y_4$ were introduced in \cite{Cattoen, Aviles} (but we did not use their $y_3$ due to its nonmonotonic dependence on the scale factor), $y_5$ was introduced by us \cite{Semiz} and \cite{Sutherland}. For function templates, we consider eight distinct families of functions, constructed from polynomials, Pad\'{e} approximants, exponentials, etc. (Table \ref{introducingfamilies}). The Pad\'{e} approximant \cite{Pade1, Pade2, Pade3}, is defined as

\begin{equation}
\tilde{P}(y,M,N) = \frac{P_M(y_i)}{P_N(y_i)+1}
\end{equation}
where $P_M(y_i)$ denotes an $M^{\rm th}$ order polynomial without constant term. 

\begin{table} [h!]
\caption{The 8 different families used in fits. $y$ can be any one of the redshift parameters described previously ($y_0$ to $y_6$), $P_N(y)$ is the $N^{\rm th}$ order polynomial with zero constant term (except when using $y_{6}=u$), $u(y)$ is $(1+z)$ expressed in terms of $y$, $\tilde{P}$($y$, $M$, $N$) is the Pad\'{e} approximant in variable $y$ and orders $M$ \& $N$; and $c$ \& $d$ are constants. If necessary, the constant term of polynomial is adjusted to make $d_L(z=0)=0$}
\centering
\begin{tabular}{| c | c |}
\multicolumn{1}{r}{}\\
\hline
Designation & Function family  \\
\hline
F1 & $P_N(y)$  \\
\hline
F2 & $P_N(y) u(y)$  \\
\hline
F3 & $P_N(y) \exp(c y)$   \\
\hline
F4 & $P_N(y) u(y) \exp(c y) $   \\
\hline
F5 & $P_N(y) \exp(c y + d y^2)$   \\
\hline
F6 & $P_N(y) u(y) \exp(c y + d y^2) $   \\
\hline
F7 &$\tilde{P}$($y$, $M$, $N$)   \\
\hline
F8 & $u(y)$ $\tilde{P}$($y$, $M$, $N$)   \\
\hline
\end{tabular}
\label{introducingfamilies}
\end{table} 

Note that in finding the best fit for a given combination of function family and redshift variable, we also leave the number of parameters free; we believe that this is one of the unique features of our LDF procedure. In this work, we select the best fit in each combination by minimizing the Bayesian Information Criterion (BIC), ensuring an optimal balance between goodness-of-fit and model complexity. The BIC method avoids overfitting by penalizing large number of parameters in the fit, appropriate here, given the large number of available data points in the Pantheon+ sample. 

\begin{landscape}
\begin{figure}[h!]
\centering
\includegraphics[width=1 \columnwidth]{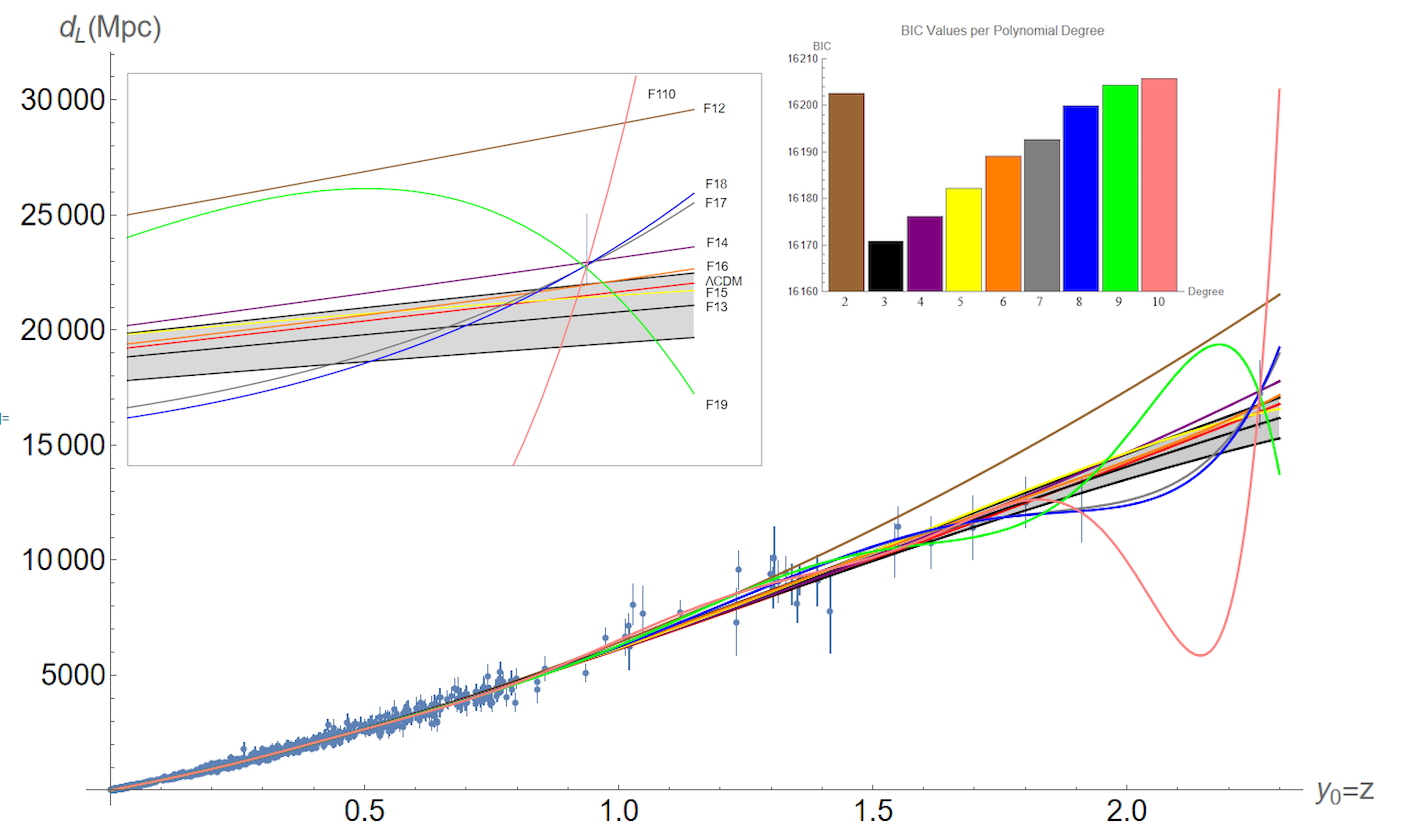}
\caption{Pantheon+ data in terms of luminosity distance and standard redshift, the $N=2$ to 10 fits for the first family, simple polynomials and ${\rm \Lambda}$CDM (red) model, and the $1$-$\sigma$ confidence-level of the best-fitting member ($N=3$) of the family F1. The insets show (i) a magnified view of the right end of the data, and (ii) a bar chart of the BIC values for the F1 fits ($N=2$ to $10$).}
\label{Y0fits only} 
\thispagestyle{empty}
\end{figure}
\end{landscape}

As a demonstration, Figure \ref{Y0fits only} displays the best polynomial fits ranging from $N = 2$ to 10  parameters, together with the BIC values; to Pantheon+ data as function of the standard redshift $z$. The figure highlights the best-fitting model within this family ($N = 3$) along with its $1$-$\sigma$ confidence region. For comparison, the theoretical ${\rm \Lambda}$CDM curve is also shown, which can be seen to lie within the $1$-$\sigma$ confidence region of the best-fitting (cubic) polynomial. The increasingly erratic behavior observed in the higher-order fits is a typical manifestation of over-fitting, in which the model starts to track individual data fluctuations rather than capturing the smooth underlying trend.

\subsection{Spread-LDF of Pantheon+ Data}

\begin{figure}[h!]
\centering
\includegraphics[width=1 \columnwidth]{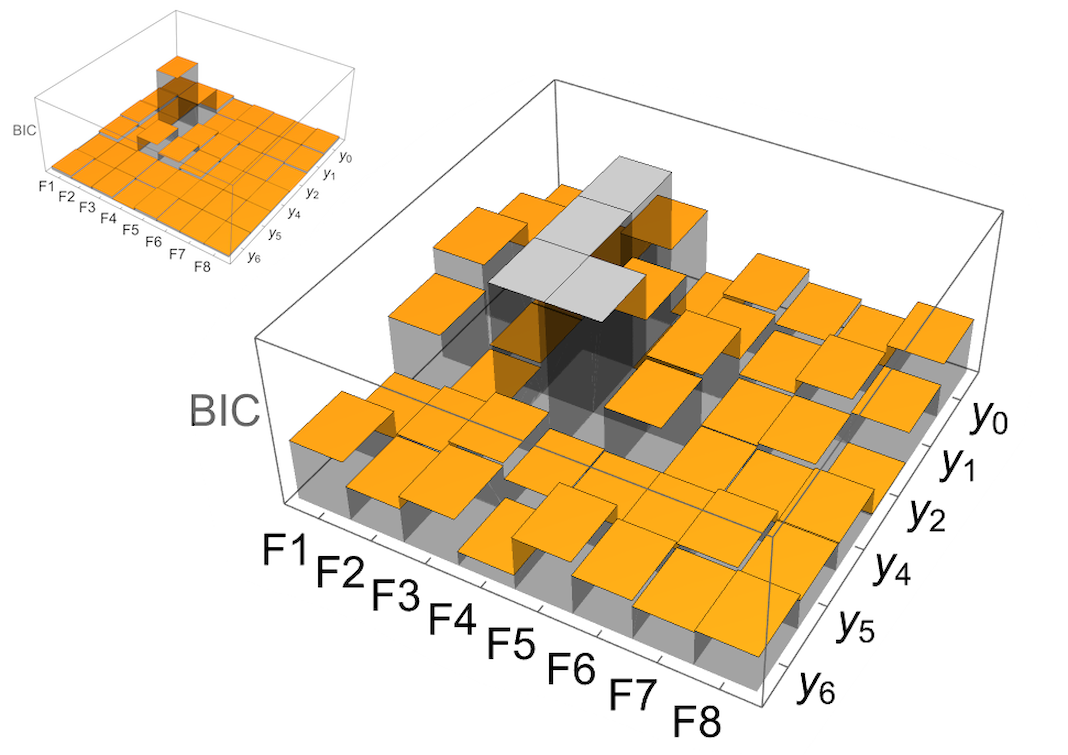}
\caption{BIC values in Table \ref{families}, illustrated as a 3D bar chart for different function families and redshift variables. The main plot (right) provides a close-up view by truncating the tallest bars for better visibility of the lower values. An inset (left) shows the full-scale chart including all bar heights for completeness.}
\label{GRB 3D bars} 
\end{figure}

All of the best fits of each family and redshift variable, together with their BIC values are shown in Table \ref{families} and illustrated in Figure \ref{GRB 3D bars}. Among all the fits, the combination of the $y_1$ redshift parameterization with the Family 4 function form (polynomial multiplied by $(1 + z)$ and exponential) yielded the lowest BIC value. This indicates that, within the set of candidate functions considered, the $y_1$-F4 combination provides the most statistically favored description of the luminosity distance data according to the BIC criterion. Nevertheless, to ensure robustness and to minimize possible biases introduced by specific functional choices, we use all fits with acceptable BIC values for further analysis. Our subsequent reconstruction of the expansion history is thus based on the ensemble of selected fits rather than solely on the single best-fit function.

\begin{landscape}
\begin{figure}[h!]
\centering
\includegraphics[width=1 \columnwidth]{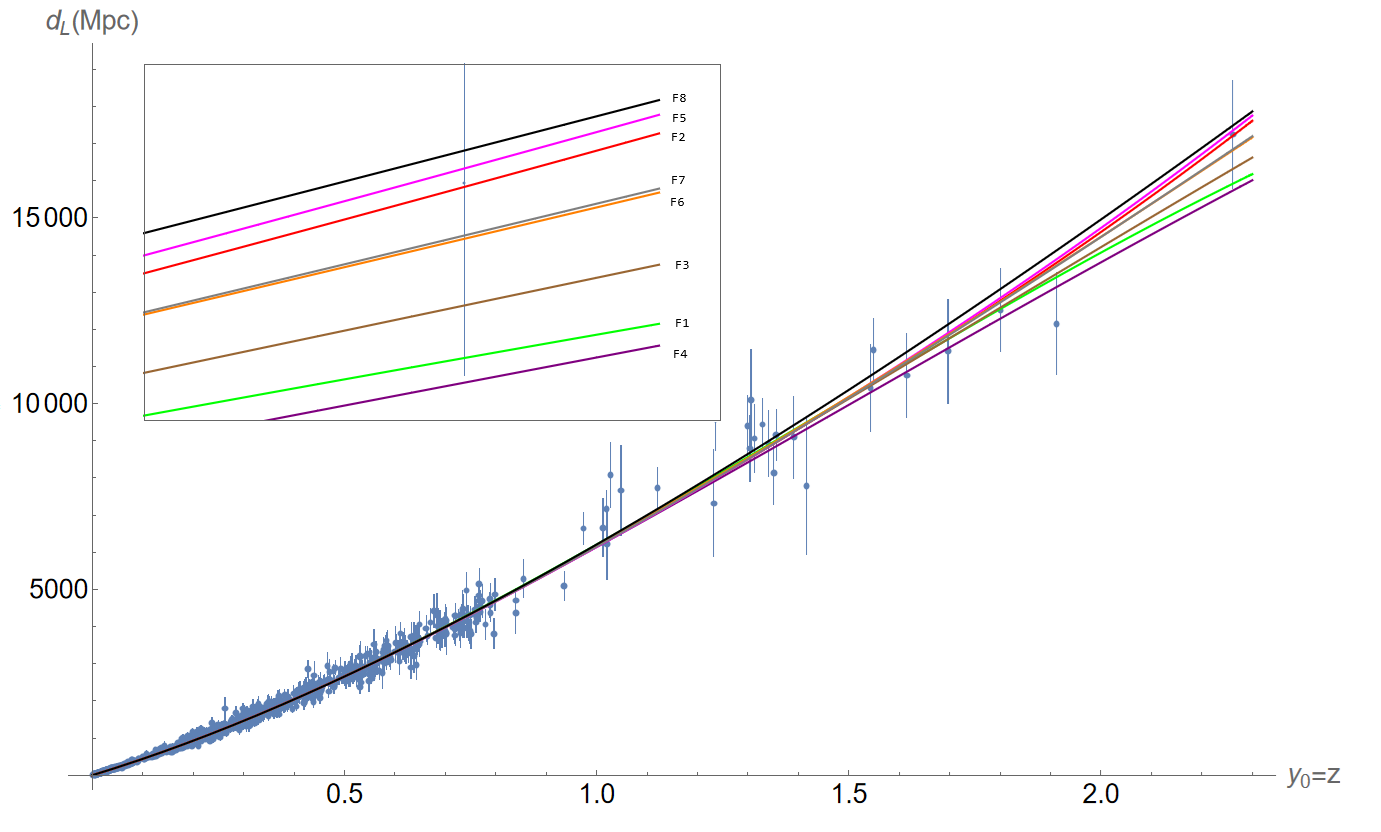}
\caption{Best-fitting curves from each functional family applied to the Pantheon+ dataset, using the standard redshift variable $y_0=z$. Each curve represents the optimal model (based on BIC) within its respective function class. This comparison illustrates how different model families approximate the observed luminosity distance–redshift relation.}
\label{Y0 all best fits} 
\thispagestyle{empty}
\end{figure}
\end{landscape}

In Figure \ref{Y0 all best fits}, we show the best-fitting model from each family of functions considered, for the standard redshift variable $y_0 = z$. The curves are plotted alongside the Pantheon+ dataset, enabling a visual comparison of how different functional forms represent the observed luminosity distance data.

\begin{table} [h!]
\caption{The best fits for the Pantheon+ and GRB data. Each cell displays the internal label(s) of the best-fitting member of the row's family for the column's redshift variable, and the fit's BIC value.}
\centering
\resizebox{14cm}{!}{
\begin{tabular}{| c | c | c | c | c | c | c |}
\multicolumn{1}{r}{}\\
\hline
\backslashbox{family}{variable}& $y_{0}=z$ & $y_{1}$ & $y_{2}$ & $y_{4}$ & $y_{5}$ & $y_{6}=u$  \\
\hline
F1 & 4; 19418.9 &  4; 19425.6 & 6; 19431.1 & 4; 19425.3 & 3; 19412.4 & 4; 19418.9 \\
\hline
F2 & 3; 19413.3 & 3; 19412.6& 3; 19412.8 & 3; 19411.8 & 3; 19413.7 & 3; 19413.3 \\
\hline
F3 & 4; 19423.6 &  2; 19516.5 & 2; 19690.4 & 2; 19480.3 & 4; 19416.8 & 3;19416.2 \\
\hline
F4 & 2; 19408.8 &  2; 19405.5 & 2; 19431.3 & 2; 19444.4 & 3; 19413.1 & 2; 19408.8 \\
\hline
F5 & 4; 19417.4 & 3;  19416.5& 4; 19423.5 & 4; 19422.7 & 3; 19412.7 & 4; 19417.4 \\
\hline
F6 & 3; 19414.4 & 3; 19412.9 & 3; 19412. & 3; 19411.8 & 3; 19412.7 & 3; 19414.4 \\
\hline
F7 & 2;1; 19413.1 & 2;1; 19416.2 & 1;2; 19412.6 & 1;1; 19411.2 & 2;1; 19414.0& 2;1; 19413.1 \\
\hline
F8 & 2;1; 19418.0 & 2;1; 19413.1 & 1;1; 19406.0 & 1;2; 19411.8 & 1;2; 19412.1 & 1;2; 19413.5 \\
\hline
\end{tabular}
}
\label{families}
\end{table}

\section{Results for Expansion and Acceleration History of the Universe}

In this section, we present and discuss the results of the procedure (spread-LDF) using the Pantheon+ data, both discussed in the previous section. This constitutes an update of the relevant part of our previous work, needed due to availability of newer data. 

\subsection{Reconstruction of the Acceleration Functions}

\begin{figure}[h!]
\centering
\includegraphics[width=1 \columnwidth]{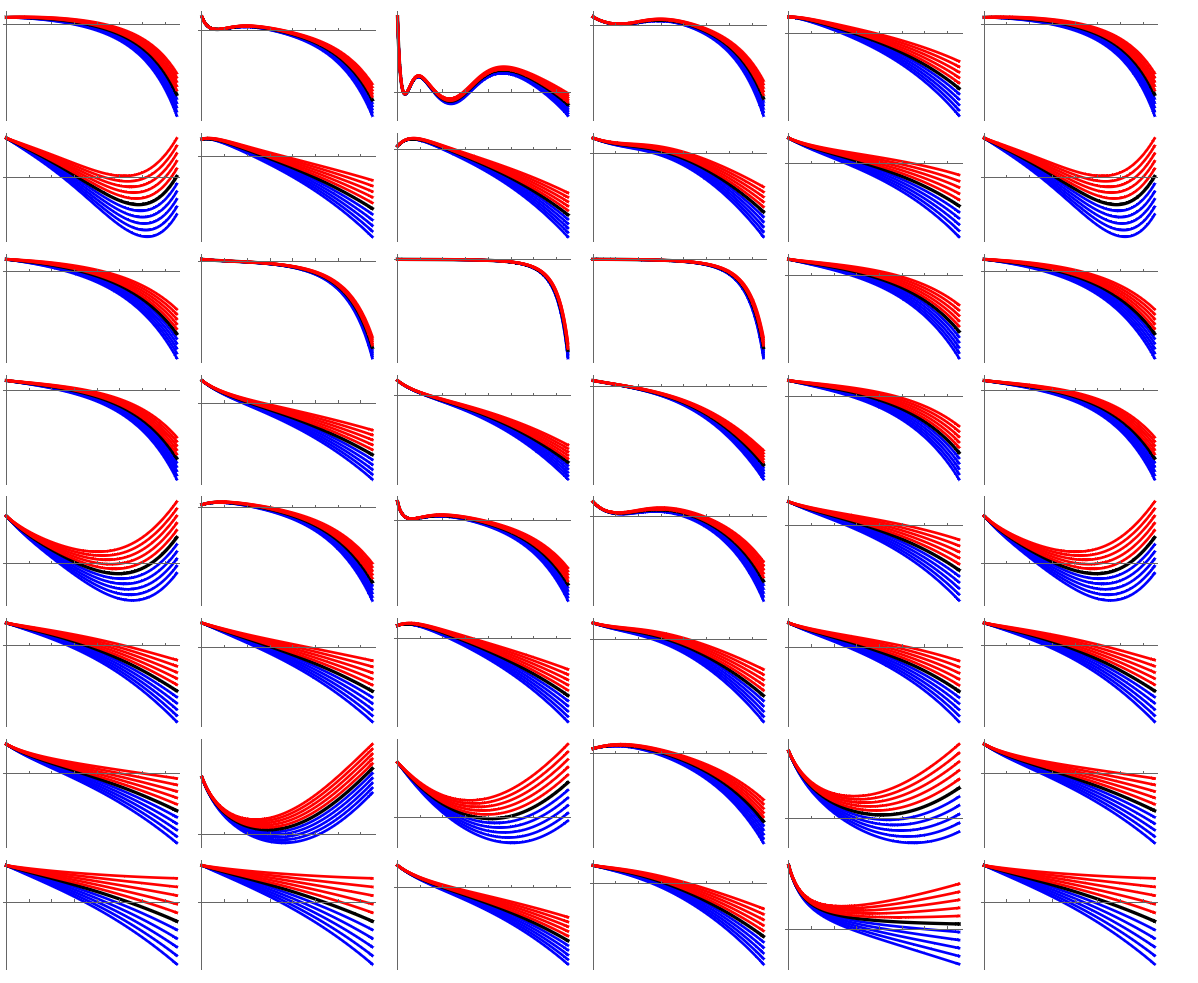}
\caption{The $\ddot{a}$ functions, computed analytically for the Pantheon+ data using Eq. (\ref{addot}). The columns represents $y_0$ to $y_6$ and rows are for families F1 to F8. Blue, black and red curves are for open, flat and closed spaces respectively, for -$\kappa_0 \leq \kappa \leq \kappa_0$. All horizontal axes converted to $y_0 = z$, with the redshift range spanning $0 \leq z \leq 1.5$.} 
\label{grid without GRBs} 
\end{figure}

Figure \ref{grid without GRBs} displays acceleration functions over the redshift range $0 < z < 1.5$, reconstructed from all 48 best-fit $d_L(z)$ functions found in Sect.2 (Table  \ref{families}) from the Pantheon+ data. Each panel corresponds to a different combination of fitting family and redshift reparameterization, but transformed to be expressed in terms of the standard redshift $z$. It can be seen that some of the reconstructed acceleration functions exhibit significant oscillatory behavior. However, these oscillations are not consistent across the different curve sets, so we decide that they are primarily artifacts arising from noise amplification during second-order differentiation and do not reflect physically meaningful features. Furthermore, some panels feature positive acceleration for large $z$, which we suspect is also an artifact.

\begin{figure}[h!]
\centering
\includegraphics[width=1 \columnwidth]{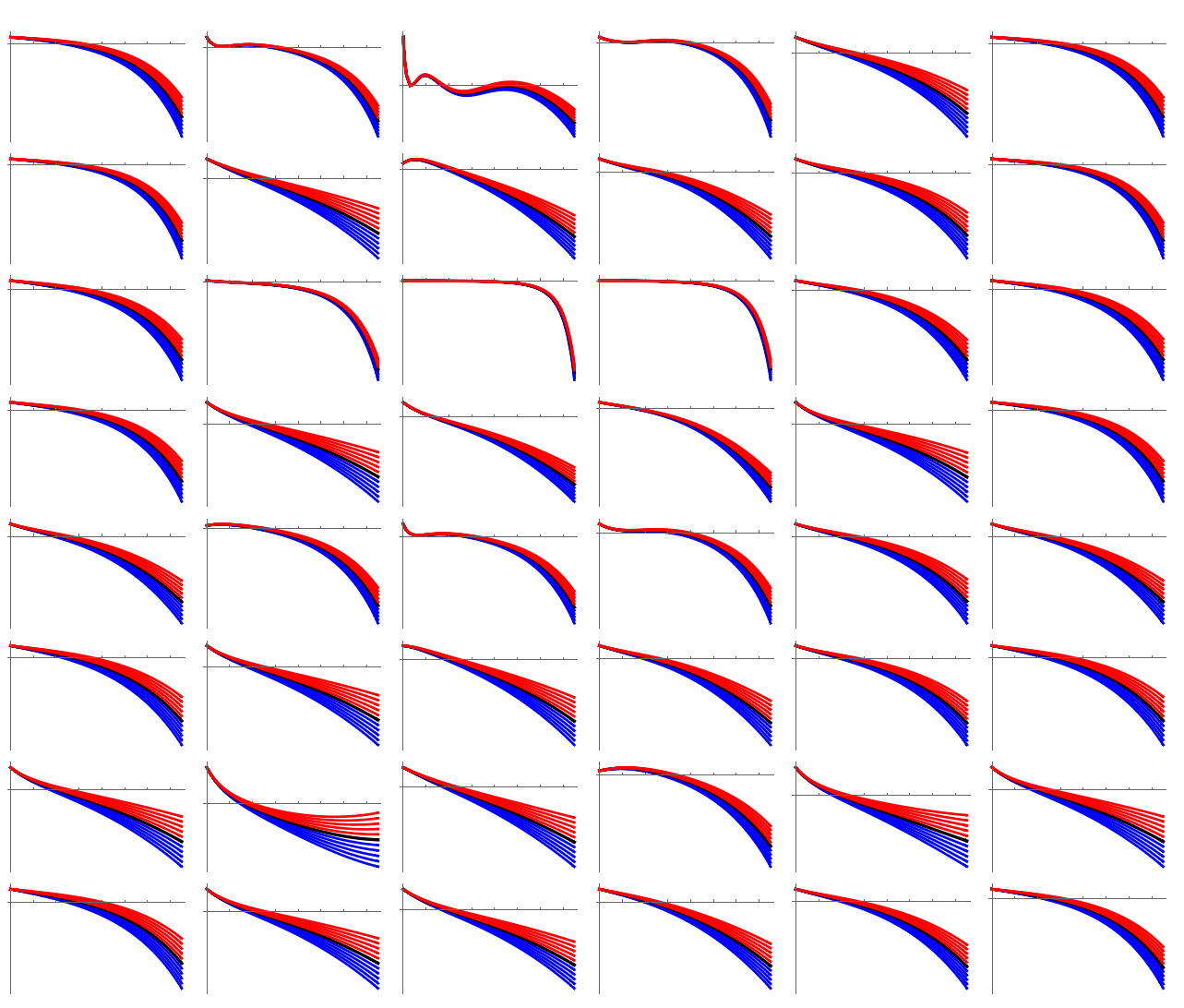}
\caption{The $\ddot{a}$ functions, computed analytically for the Pantheon+ and GRB data using Eq. (\ref{addot}). The columns represents $y_0$ to $y_6$ and rows are for families F1 to F8. Color encodes curvature as in Figure \ref{grid without GRBs}. All horizontal axes converted to $y_0 = z$, with the redshift range spanning $0 \leq z \leq 1.5$.} 
\label{GRID with GRBs} 
\end{figure}

As discussed in Sections 1 and 2, we include the GRB data in an effort to tame these artifacts, after rescaling as described in Section 2.1. The results are shown in Fig.\ref{GRID with GRBs}, showing that the GRB data primarily act to stabilize the high-redshift reconstructions while leaving the low-redshift behavior, which is most critical for determining the transition from deceleration to acceleration, largely unchanged.

\subsection{Selection of Physically Acceptable Reconstructions}

Despite the stabilization due to inclusion of the GRB data, some inconsistent anomalous behavior remains in the reconstructed acceleration graphs. To systematically select acceptable reconstructions, we applied an elimination procedure based on the grid layout of the 48 functions, that is, rows (fitting family) and columns (redshift reparameterization). Entire rows or columns, where oscillatory or unphysical behavior was dominant in one figure, were eliminated from consideration. This conservative approach ensures that the selection criteria are applied uniformly across all function families and parameterizations, minimizing subjective bias.

In particular, we eliminated:
\begin{itemize}
\item{Rows 1 and 5, corresponding to F1 and F5 (from table \ref{introducingfamilies})},
\item{Columns 3 and 4, corresponding to $y_2=\textnormal{arctan}\frac{z}{1+z}$, $y_4=\textnormal{arctan}\; z$}
\end{itemize}
This leaves us with 24 reconstructions to continue with.

\subsection{Reconstructed Expansion: Acceleration and Hubble Parameter}

\begin{figure}[h!]
\centering
\includegraphics[width=1 \columnwidth]{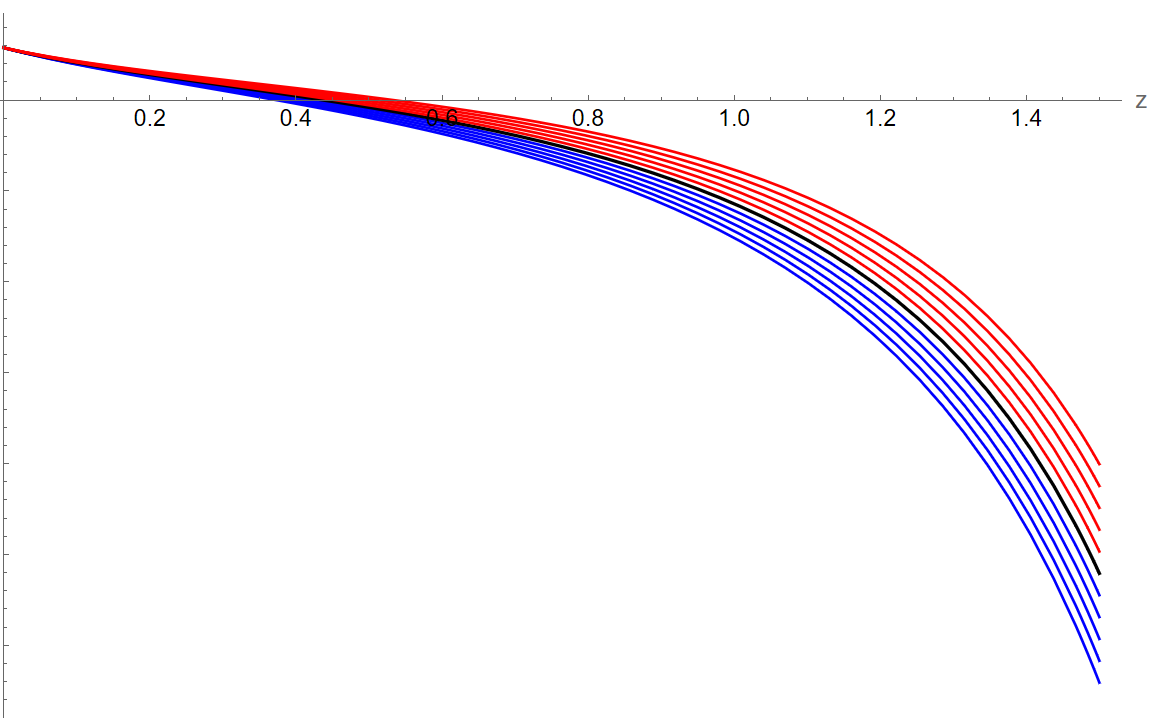}
\caption{Reconstructed average acceleration, $\ddot{a}(z)$. Color encodes curvature as in Figures \ref{grid without GRBs} \& \ref{GRID with GRBs}.} 
\label{average_acceleration_all_curves} 
\end{figure}

We now proceed to averaged acceleration curves. For each of the 11 acceleration values, the average $\ddot{a}_{\rm avg}(z)$  was constructed by simply taking the arithmetic mean of the 24 individual $\ddot{a}(z)$ functions at each redshift point (Figure \ref{average_acceleration_all_curves}). It is evident that at $z = 0$, all remaining acceleration functions, as well as the averaged curve, yield $\ddot{a}(0) > 0$, confirming the current accelerated expansion of the universe. Also, all remaining functions feature a transition to deceleration as we go into the past; as expected.

\begin{figure}[h!]
\centering
\includegraphics[width=1 \columnwidth]{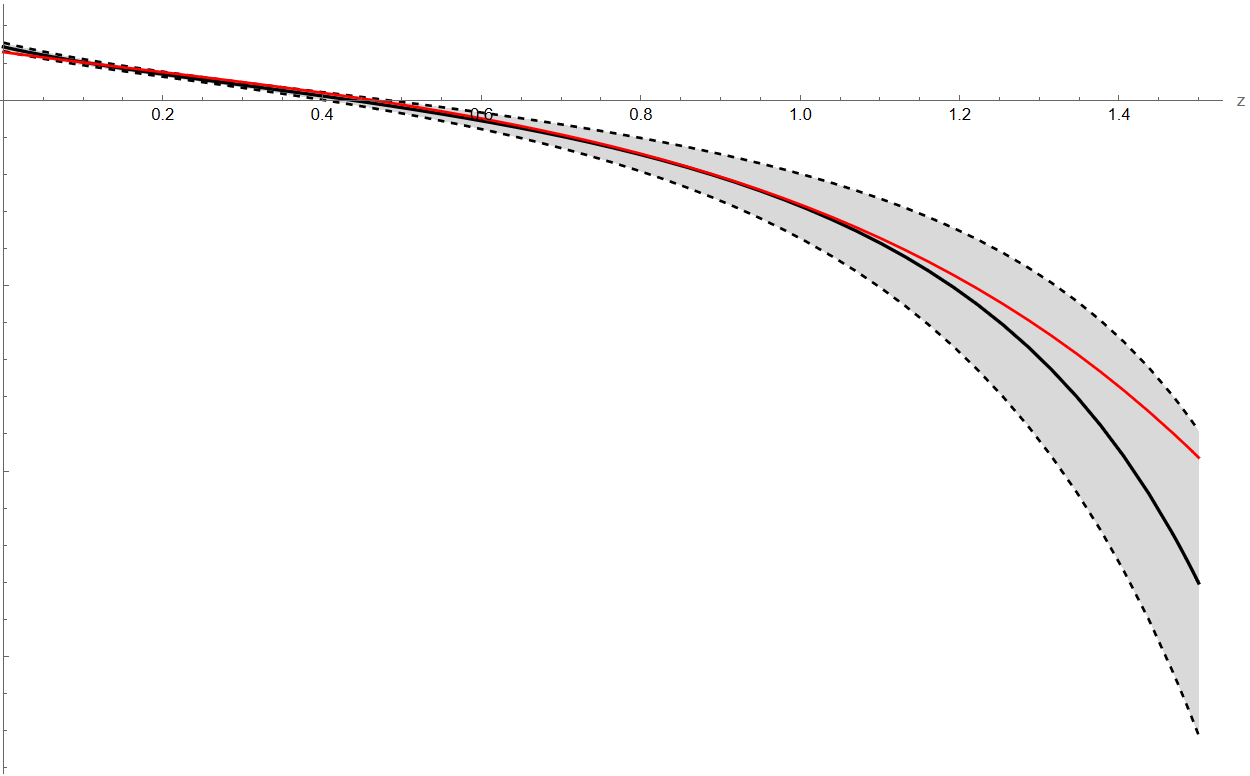}
\caption{Reconstructed average acceleration $\ddot{a}(z)$ with $1$-$\sigma$ confidence band, shown for the flat curvature case ($\kappa=0$). The shaded region reflects propagated uncertainty from the representative fit ($y_5$-F3, given in red). The transition redshift $z_T$ is the zero crossing.} 
\label{ftransition} 
\end{figure}

To estimate the uncertainty associated with the averaged curve, we selected a representative function among the 24 fits for the spatially flat case. Given the similarity in fitting behavior among the selected functions, this approximation provides a reasonable estimate of the uncertainty without significantly affecting the derived transition redshift. The representative fit was chosen as the one minimizing the integrated absolute difference with respect to the averaged curve over the redshift range $0 < z < 1.5$, that is, minimizing

\begin{equation}\label{minimize}
\Delta= \int^{1.5}_0  \left|\ddot{a}_{\textnormal{fit}}(z)-\ddot{a}_{\textnormal{avg}}(z)\right|dz.
\end{equation}
for $\kappa = 0$. With this method, fit $y_5$-F3 becomes the representative fit. The uncertainty on $\ddot{a}$ was propagated from the fit parameters of the representative function using standard linear error propagation (see Section 2.2), including covariance terms. In Figure \ref{ftransition}, we present the averaged $\ddot{a}(z)$ curve for $\kappa = 0$ together with the closest acceleration reconstructed from $y_5$-F3, and the $1$-$\sigma$ confidence band calculated for it.

For completeness, we also extract the transition redshift $z_T$ for the extreme curvature values $\kappa=\pm \kappa_0$. The results are summarized in Table \ref{transition table}, showing that while spatial curvature slightly shifts the value of $z_T$, the overall conclusion of a transition from deceleration to acceleration remains robust.
\begin{table}[h!]
\caption{Reconstructed transition redshifts for different spatial curvatures}
\centering
\begin{tabular}{l|l}
Curvature $\kappa$ &$ z_T\pm \delta z_T$ \\ \hline
     -$\kappa_0$        &  $0.37^{+0.03}_{-0.03}$ \\
         0                  & $0.44^{+0.04}_{-0.03}$ \\
     $+\kappa_0$     & $ 0.53^{+0.08}_{-0.05}$
\end{tabular}
\label{transition table}
\end{table}

\begin{figure}[h!]
\centering
\includegraphics[width=1 \columnwidth]{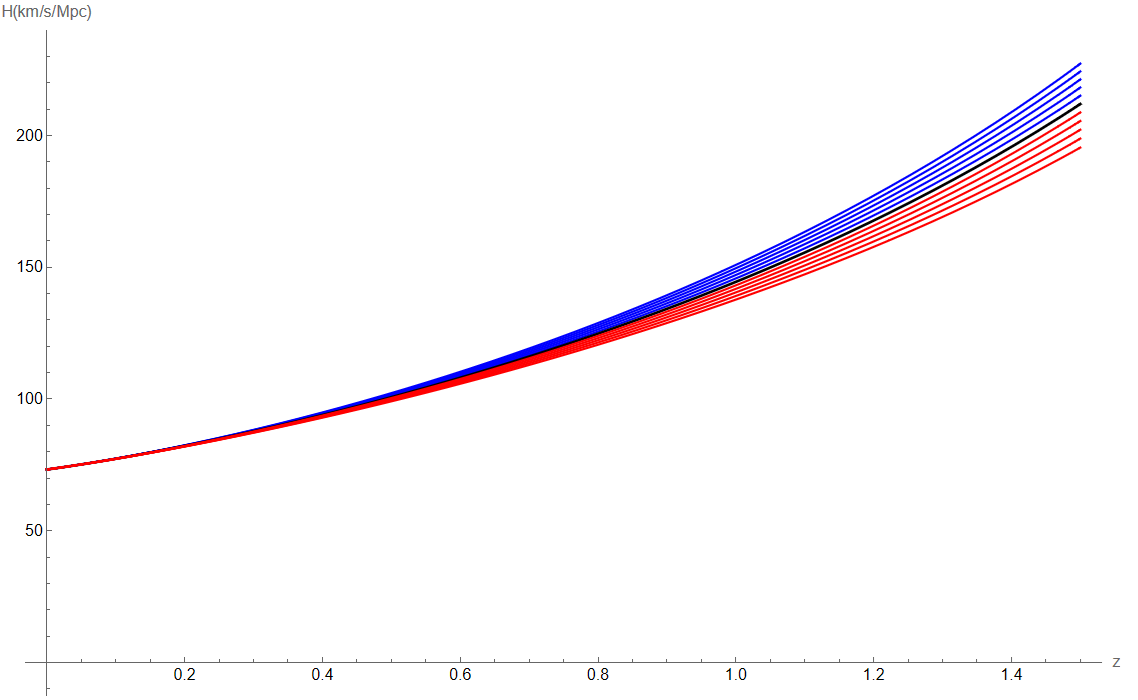}
\caption{Reconstructed Hubble parameter $H(z)$ obtained from the average of 24 accepted $\dot{a}(z)$ functions. Color encodes curvature as in Figures \ref{grid without GRBs}-\ref{average_acceleration_all_curves}. The present-day value for the flat case is $H_0 \approx73.2$ km/s/Mpc.} 
\label{Hubble average} 
\end{figure}

Using the average of the reconstructed $\dot{a}(z)$ functions, we can compute the Hubble parameter via Eq. (\ref{hubble}) and we show the resulting $H(z)$ curves corresponding to the 11 values of the curvature parameter $\kappa$ in Figure \ref{Hubble average}. The flat case ($\kappa=0$) yields a present-day Hubble constant of
\begin{equation}\label{hubble0reconstructed}
H_0=H(z=0)\approx 73.23 \pm 0.24 \; \textnormal{km/s/Mpc},
\end{equation}
which is in agreement with local measurements from SH0ES \cite{shoes}. However, this result should not be interpreted as an independent constraint on $H_0$, but rather as a consistency check within a SH0ES-calibrated framework, since the Pantheon+ distance moduli are calibrated using a fiducial absolute magnitude $M$ derived from SH0ES 2021 Cepheid host distances, and our reconstructed $H(z)$ inherits this normalization. The quoted error does not reflect full cosmological uncertainty either, as it omits correlated systematics. The increasing trend of $H(z)$ with redshift is consistent with expectations from standard cosmological models, where the expansion rate was higher in the past due to matter domination. The spread among the curves for different $\kappa$ values illustrates the degeneracy between spatial curvature and the expansion history, where different values of $\kappa$ can lead to similarly shaped $H(z)$ curves, particularly at low redshifts.

\begin{figure}[h!]
\centering
\includegraphics[width=1 \columnwidth]{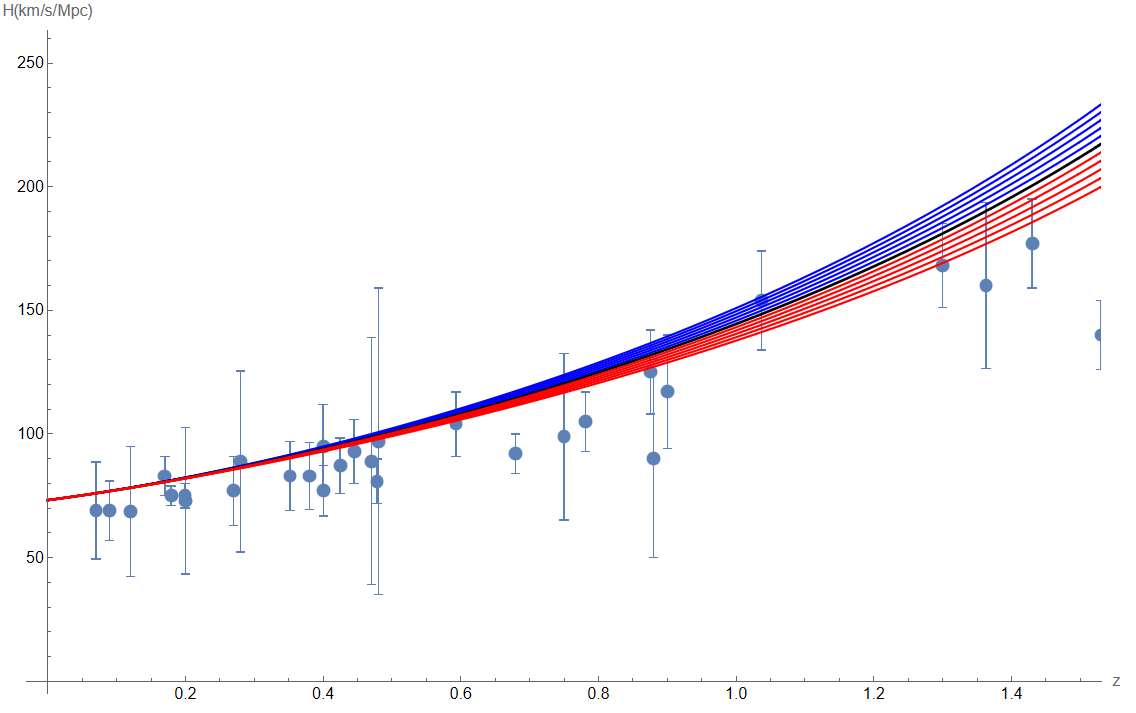}
\caption{Comparison between the reconstructed Hubble parameter $H(z)$ (Color encodes curvature as in Figures \ref{grid without GRBs}-\ref{average_acceleration_all_curves} \& \ref{Hubble average}) and observational data from cosmic chronometers \cite{Moresco}.} 
\label{CCM} 
\end{figure}

As an additional consistency check, we compare our reconstructed $H(z)$ to direct Hubble parameter measurements from cosmic chronometers \cite{Moresco}. As shown in Figure \ref{CCM}, our model-independent curves remain within the $1$-$\sigma$ range of the observational data across $0<z<1.5$, demonstrating good agreement and further validating the reliability of the reconstruction.

\section{Semi-Model-Independent Analysis: Dark Energy with Positive Pressure}

Having completed our model-independent reconstruction of the expansion history in previous section, we now transition to a semi-model-independent analysis; that is, we assume a gravity model (GR) but no matter/energy model. So, we use the Einstein field equations to interpret the reconstructed quantities in terms of stress-energy-momentum tensor components. This step allows us to compare the reconstructed expansion history with standard cosmological models such as ${\rm \Lambda}$CDM, and evaluate the inferred matter and dark energy content; still with the underlying assumptions about geometry, which ultimately follow from the cosmological principle.

In several parts of this section, we treat reconstructed functions such as $H(z)$, $\rho(z)$, and $w(z)$ as effective datasets in order to fit semi-model-dependent expressions. These functions are evaluated at discrete redshift intervals, and their uncertainties are estimated using the $1$-$\sigma$ error propagation from the representative fit $y_5$–F3. This approach avoids the prohibitively complex task of propagating the full set of uncertainties from all 24 accepted fits, each of which involves several parameters. While the $y_5$–F3-derived uncertainty band provides a reasonable approximation we note that this method neglects correlations between adjacent redshift points that naturally arise from shared functional dependencies. As such, the resulting fits should be interpreted as indicative rather than precision estimates.

\subsection{Interpreting the Reconstructed Expansion: Energy Density}

\begin{figure}[H]
\centering
\includegraphics[width=1 \columnwidth]{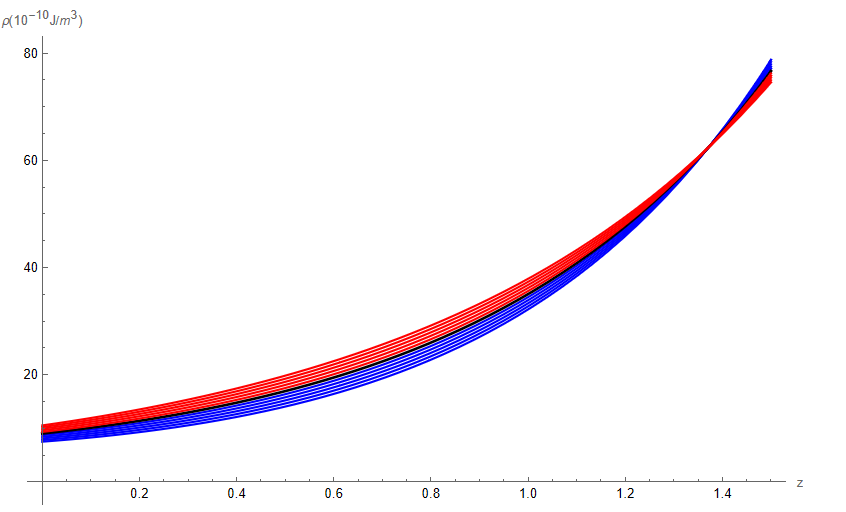}
\caption{Reconstructed total energy density $\rho(z)$ for 11 values of the curvature parameter $\kappa$, computed using the representative fit ($y_5$–F3). Color encodes curvature as in Figures \ref{grid without GRBs}-\ref{average_acceleration_all_curves}, \ref{Hubble average} \& \ref{CCM}.} 
\label{rho} 
\end{figure}

Now we can derive the total energy density of the universe as function of redshift, $\rho(z)$, within the framework of General Relativity from the first Friedmann equation and the reconstructed $\dot{a}(z)$:
\begin{equation}
\rho(z)=\frac{3c^2}{8 \pi G}\left(H^2(z)+\kappa c^2 (1+z)^2\right)
\end{equation}
The resulting energy density functions are shown in Figure \ref{rho}, again for the range of curvatures considered in this work. We believe that the point where the reconstructed energy density is independent of spatial curvature does not have a cosmological significance.

\begin{figure}[h!]
\centering
\includegraphics[width=1 \columnwidth]{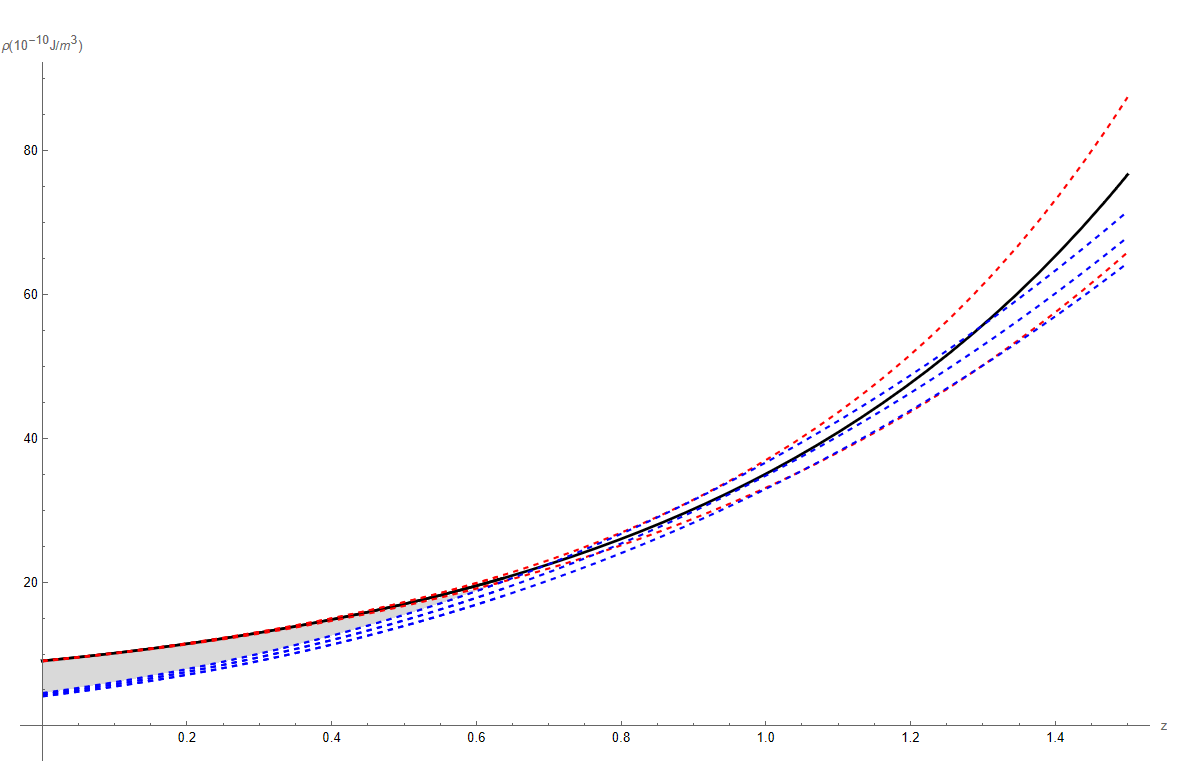}
\caption{Flat curvature case: total energy density $\rho(z)$(black), with $1$-$\sigma$ upper and lower bounds (dashed-red curves) propagated from the representative fit. Each is compared with a scaled matter-only evolution $\rho_m(z)=\rho_{m,0}(1+z)^3$ (dashed-blue curves), matched at $z\approx 1$. The shaded region between $\rho(z)$ and the matter curve at low redshift highlights the minimum extra energy density required beyond matter.} 
\label{flat rho} 
\end{figure}

We also compute a representative $1$-$\sigma$ confidence band for $\rho(z)$ by propagating the uncertainty from the representative fit function $y_5$-F3 (see Section 3.3), and apply it to the averaged flat case. This yields three curves: $\rho(z)$, $\rho(z)+\sigma_{\rho}(z)$, and $\rho(z)-\sigma_{\rho}(z)$ shown in Figure \ref{flat rho}, the latter two shown in dashed red.

   We do know that matter exists in the universe, and its density is of the form
\begin{equation}\label{rho_matter_evolution}
\rho_m(z)=\rho_{m,0}(1+z)^3,
\end{equation}
To represent the maximum amount of matter our universe can accommodate, we adjust the parameter $\rho_{m,0}$  as large as possible, but with the matter density curve staying always below the total energy density curve. Applying this to all three curves mentioned above, we get the three dashed blue curves showing the maximum possible matter amount at any $z$, and its representative confidence interval, for the spatially flat case. The shaded region then necessarily represents the minimum amount of the non-matter component—consistent with dark energy.

At redshift $z=0$, we find the reconstructed total energy density to be
\begin{equation}
\rho(0)=9.04 \pm 0.06 \;\textnormal{in units of $10^{-10}$Joule/m$^3$}.
\end{equation}
The corresponding scaled matter-only curve yields
\begin{equation}
\rho_{m, {\rm max}}(0)=4.35\pm 0.23 \;\textnormal{in units of $10^{-10}$Joule/m$^3$}
\end{equation}
This leads to a clear excess of energy density at low redshift:
\begin{equation}
\rho_{\textnormal{DE,min}}(0)=\rho(0)-\rho_m(0)=4.69\pm0.24 \;\textnormal{in units of $10^{-10}$Joule/m$^3$}
\end{equation}
which can be interpreted as a lower bound on the dark energy density under the assumption of flat geometry and General Relativity. These values correspond to the following fractional density parameters:
\begin{equation}
{\rm \Omega}_m<\frac{\rho_m(0)}{\rho(0)}=0.481^{+0.029}_{-0.028}, \quad {\rm \Omega}_\Lambda>\frac{\rho_\Lambda(0)}{\rho(0)}=0.519^{+0.028}_{-0.029}.
\end{equation}
These values reinforce the conclusion that, even without any assumptions on the contents of the universe (beyond the default existence of matter/dust to some unspecified extent), the expansion history implied by the data requires a non-matter component to explain the observed acceleration — which can be called dark energy. Recall that the conclusion is reached under assumptions of the Cosmological Principle and that General Relativity remains valid on cosmological scales.

\subsection{${\rm \Lambda}$CDM Fit to Combined Distance Data and Reconstructed $H(z)$}

To compare our model-independent reconstruction with standard cosmology, we fit a flat ${\rm \Lambda}$CDM model to the combined Pantheon+ and GRB luminosity distance dataset; and then we fit a flat ${\rm \Lambda}$CDM Hubble function to our reconstructed $H(z)$, again for $\kappa=0$. 

The ${\rm \Lambda}$CDM prediction for the luminosity distance is
\begin{equation}
d_L(z)=(1+z)\frac{c}{H_0}\int^z_0\frac{dz'}{\sqrt{{\rm {\rm \Omega}}_m(1+z')^3+(1-{\rm \Omega}_m)}},
\end{equation}
which we fit numerically. Fixing $H_0=70$ km/s/Mpc yields a best-fit value of ${\rm \Omega}_m=0.56$, which is higher than typical ${\rm \Lambda}$CDM values. However, allowing $H_0$ to vary freely leads to a significantly improved fit, yielding:
\begin{equation}
{\rm \Omega}_m=0.37, \quad H_0 \approx 73\; \textnormal{km/s/Mpc}
\end{equation}
which is fully consistent with our reconstructed Hubble function (Section 3.3), as well as SH0ES \cite{shoes} estimates.

The theoretical ${\rm \Lambda}$CDM prediction for $H(z)$ is given by:
\begin{equation}
H(z)=H_0\sqrt{{\rm \Omega}_m(1+z)^3+(1-{\rm \Omega}_m)},
\end{equation}
where $H_0$ is the Hubble constant and ${\rm \Omega}_m$ is the present-day matter density parameter. For a fit to this function, we construct synthetic $H(z)$ data by sampling the reconstructed $H(z)$ (derived from the average of the 24 accepted $\dot{a}(z)$ functions as described in Section 3.3) at equally spaced redshift points and assigning uncertainties based on the error band derived from the representative fit $y_5$-F3.

\begin{figure}[H]
\centering
\includegraphics[width=1 \columnwidth]{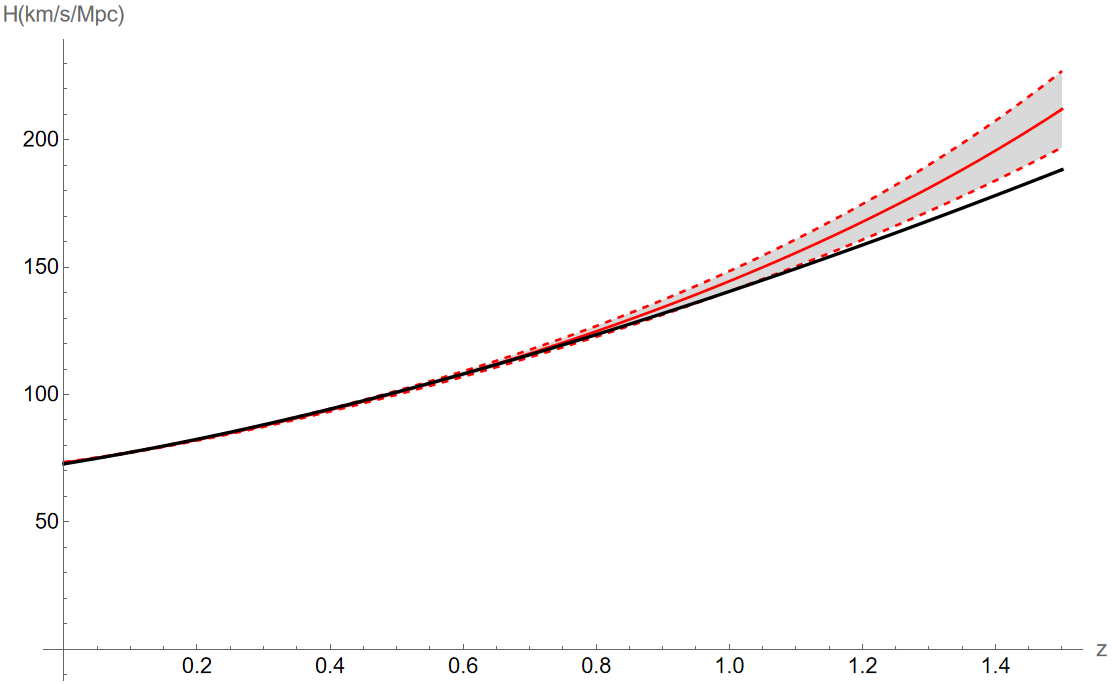}
\caption{Reconstructed Hubble parameter $H(z)$ from the average of 24 accepted $\dot{a}(z)$ curves (with $\kappa=0$), shown with a $1$-$\sigma$ error band derived from the representative $y_5$-F3 function. The solid black curve shows the best-fit flat ${\rm \Lambda}$CDM model, yielding $H_0\approx72.8$ km/s/Mpc, ${\rm \Omega}_m\approx0.39$.} 
\label{LCDM_fit_to_H} 
\end{figure}

Using these uncertainties in a weighted least-squares fit, we find:
\begin{equation}
{\rm \Omega}_m\approx0.39, \quad H_0 \approx 72.8 \textnormal{ km/s/Mpc},
\end{equation}
which are consistent with local distance ladder measurements of $H_0$ (e.g., SH0ES) subject to the same caveat as associated with (\ref{hubble0reconstructed}), though the matter density is slightly elevated compared to Planck’s \cite{Planck} value of ${\rm \Omega}_m \approx 0.31$. The resulting ${\rm \Lambda}$CDM curve is shown in Figure \ref{LCDM_fit_to_H}, overlaid on the reconstructed $H(z)$ with $1$-$\sigma$ confidence bands. 

These best-fit values are in close agreement with those obtained earlier in this section by fitting the ${\rm \Lambda}$CDM luminosity distance expression directly to the combined Pantheon+ and GRB data. Although the two approaches — one model-dependent, the other model-independent — are conceptually distinct, they yield consistent estimates of $H_0$ and ${\rm \Omega}_m$ within the redshift range where observational constraints are strongest. This consistency supports the reliability of our reconstruction framework and suggests that the standard ${\rm \Lambda}$CDM model remains broadly compatible with the reconstructed expansion history, at least up to $z<1$.

\subsection{Effective Equation of State from Reconstructed Dynamics}

In addition to reconstructing the total energy density $\rho(z)$, we now derive the effective pressure $p(z)$ using the second Friedmann equation:
\begin{equation}\label{frw pressure}
p(z)=\frac{c^2}{8 \pi G}\left(-\kappa c^2 (1+z)^2-H(z)^2-2\frac{\ddot{a}(z)}{a_0}(1+z)\right).
\end{equation}
This expression allows us to compute the pressure evolution of the cosmic fluid under the assumption of spatial homogeneity and General Relativity. Using the representative reconstruction ($y_5$-F3), we evaluate both $\ddot{a}$ and $\rho(z)$ for the flat curvature case $\kappa=0$, and compute the effective equation of state parameter:
\begin{equation}\label{simple EoS}
w(z)=\frac{p(z)}{\rho(z)}.
\end{equation}
This function describes the total dynamical behavior of the universe's energy content as a function of redshift, without assuming any particular decomposition into matter and dark energy components.

\begin{figure}[h!]
\centering
\includegraphics[width=1 \columnwidth]{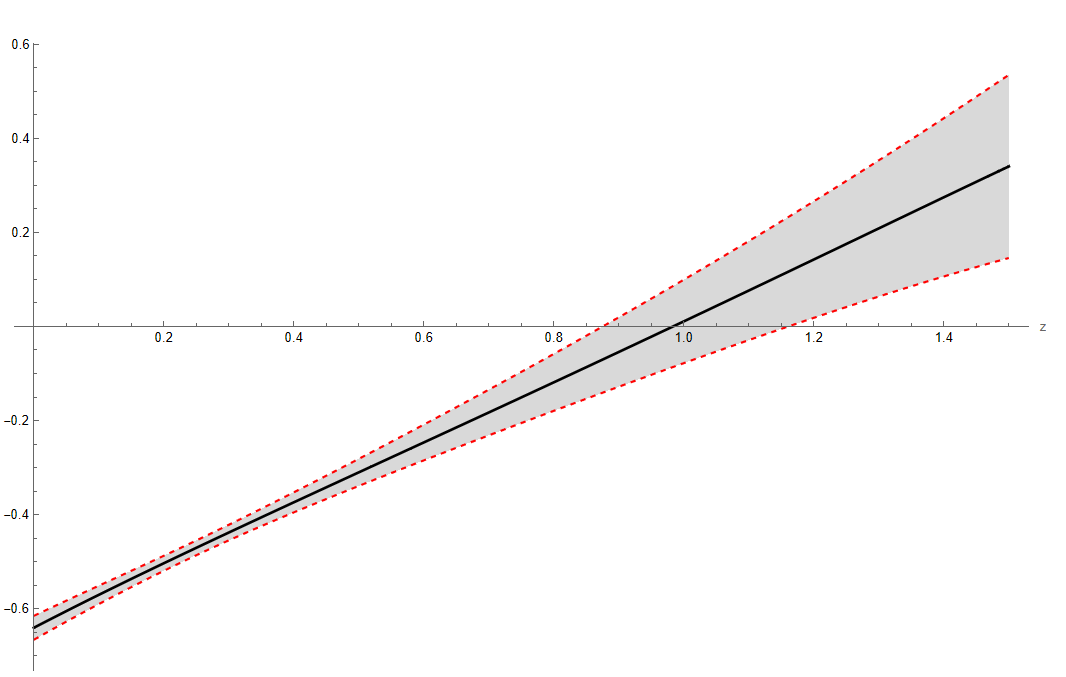}
\caption{Reconstructed effective equation of state parameter $w(z)=p(z)/\rho(z)$ using the representative function ($y_5$–F3) for the flat curvature case. The present-day value is $w(0)=-0.64$, indicating accelerated expansion. At higher redshifts, $w(z)$ increases toward 0, consistent with a matter-dominated universe.} 
\label{EoSwithErrors} 
\end{figure}

In Figure \ref{EoSwithErrors} , we plot the reconstructed $w(z)$ over the redshift range $0<z<1.5$. At redshift $z=0$, we find
\begin{equation}
w(0)=-0.640^{+0.025}_{-0.025} \quad (\textnormal{for}\; \kappa=0)
\end{equation}
indicating that the effective cosmic fluid behaves as a mixture of pressureless matter and a dark energy–like component with negative pressure. The value lies between the matter-like value $w=0$ and the cosmological constant value $w=-1$, consistent with a transitioning universe. 

\subsubsection{Linearity of the Reconstructed $w(z)$}

\begin{figure}[h!]
\centering
\includegraphics[width=1 \columnwidth]{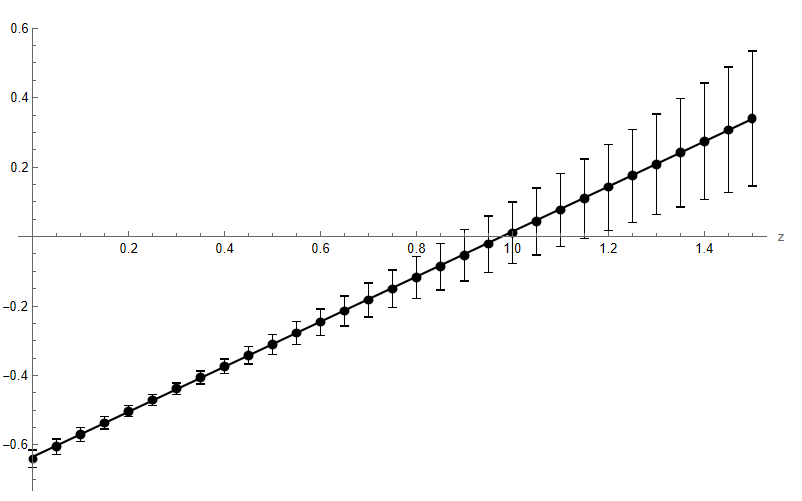}
\caption{Reconstructed effective equation of state $w(z)$ using the representative fit ($y_5$-F3) for the flat case. The dots shows the reconstructed $w(z)$; the black line is the best-fit linear model $w(z)\approx-0.63+0.65z$. The near-linearity supports the use of low-order parametrizations for describing the equation of state in this redshift range.} 
\label{fit_to_Eos} 
\end{figure}

Upon inspecting the reconstructed effective equation of state function $w(z)=p(z)/\rho(z)$, we observe that it exhibits a remarkably linear behavior over the redshift interval $0<z<1.5$. To quantify this observation, we perform a quadratic fit to the reconstructed $w(z)$ in this interval. For this purpose, we construct synthetic $w(z)$ data as was done for $H(z)$ in Sect. 4.2, again by sampling equally spaced values and using error estimates based on the  fit $y_5$–F3 (Figure \ref{fit_to_Eos}).  We do not make a Taylor expansion of the analytically "known" function $H(z)$ for two reasons, the first being that a Taylor expansion will also be influenced by the behaviour of the function beyond the range, where it is not really representative of our data; and the second, that it will not allow incorporation of the errors which are not uniformly spread along $z$. The weighted least-squares fit yields
\begin{equation}
w(z)\approx -0.64 +0,66z-0.01z^2
\end{equation}
where the smallness of the quadratic term confirms that $w(z)$ is well approximated by a linear form across the redshift range considered. If we then default to a linear fit using same synthetic data and errors (weights), we get
\begin{equation}
w(z)\approx -0.63 + 0.65z.
\end{equation}

It is important to emphasize that the $w(z)$ reconstructed here represents the effective equation of state of the total cosmic fluid, derived directly from the total energy density and pressure as inferred from the expansion history. It does not rely on any explicit separation between matter and dark energy components. We also note that the computation of $w(z)$ relies on both $\dot{a}(z)$ and $\ddot{a}(z)$, and is thus sensitive to small fluctuations in the input $d_L(z)$ fits, the $y_5$-F3 representative fit was selected because its resulting $\ddot{a}(z)$ curve most closely matches the average of all physically acceptable reconstructions (see Section 3.3). We therefore consider it a reliable proxy for extracting the effective equation of state.

\subsection{CPL Fit to the Reconstructed Hubble Function}

We now investigate whether a cosmological model with the somewhat popular time-varying equation of state of the Chevallier–Polarski–Linder (CPL)\cite{CPL,CPL2} form provides a better fit to the reconstructed expansion history than ${\rm \Lambda}$CDM. The CPL parametrization assumes
\begin{equation}
w_0+w_a\frac{z}{1+z}
\end{equation}
and predicts a Hubble function of the form:
\begin{equation}
H(z)=H_0\sqrt{{\rm \Omega}_m(1+z)^3+{\rm \Omega}_{{\rm CPL}} g(z)},
\end{equation}
where
\begin{equation}
g(z)=\textnormal{exp}\left[\int^z_0\frac{1+w(z')}{1+z'}dz'\right]=(1+z)^{3(1+w_0+w_a)}\textnormal{exp}\left[-3w_a\frac{z}{1+z}\right] .
\end{equation}
Here, ${\rm \Omega}_{\textnormal{CPL}}=1-{\rm \Omega}_\textnormal{m}$ ensures flatness.

We fit this model to $H(z)$ data reconstructed as described in the above subsection. The resulting best-fit parameters are:
\begin{equation}
H_0\approx 73.1\pm0.3 \; \textnormal{km/s/Mpc},\; {\rm \Omega}_m=0.47, \;w_0\approx-1.17, \; w_a \approx -1.41.
\end{equation}
These values indicate a present-day equation of state more negative than a cosmological constant, and a significant evolution toward less negative values at higher redshift.

To assess the robustness of the result, we also performed a fit with $H_0$ fixed to the reconstructed value $H(z=0)=73.2$ km/s/Mpc. This fixed-$H_0$ fit produced a slightly higher BIC value, with ${\rm \Delta}$BIC=4, favoring the floating-$H_0$ model. Additionally, the CPL model fit provides a substantially lower BIC value than the ${\rm \Lambda}$CDM fit described in Section 4.2, indicating a better overall description of the reconstructed expansion history.

\begin{figure}[H]
\centering
\includegraphics[width=1 \columnwidth]{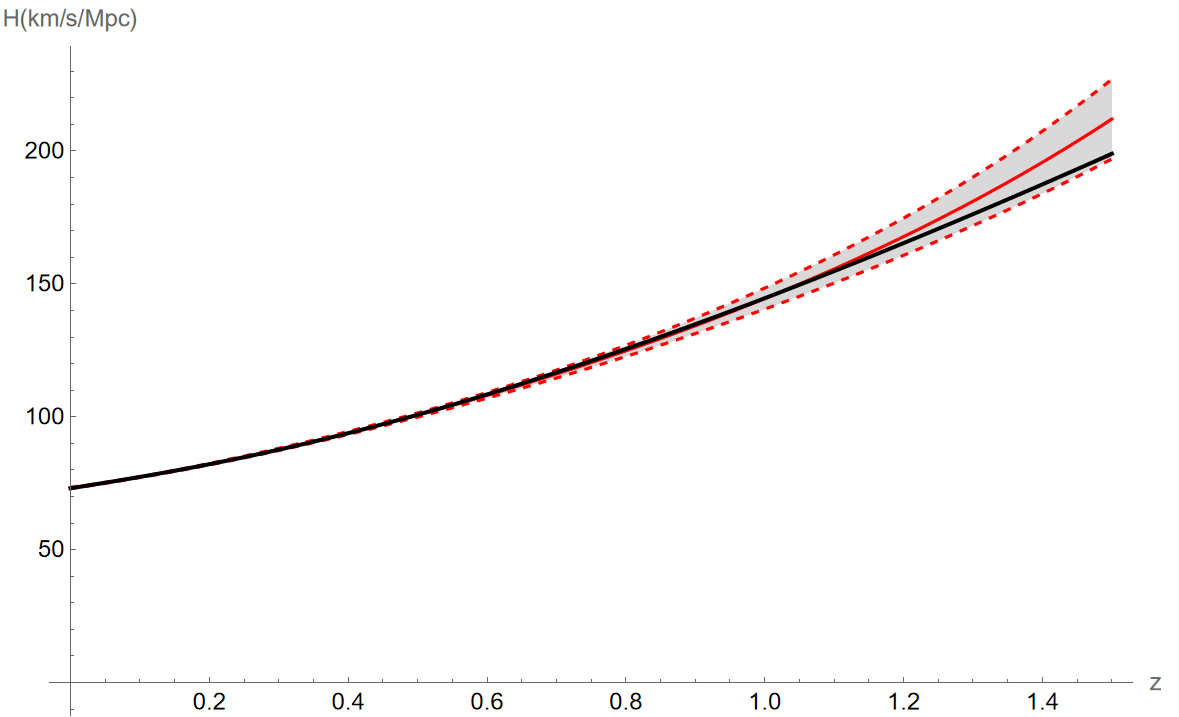}
\caption{Reconstructed Hubble parameter $H(z)$ from the average of 24 accepted $\dot{a}(z)$ functions (red), with $1$-$\sigma$ uncertainty band derived from the representative $y_5$–F3 reconstruction (dashed red curves, shaded region). The best-fit CPL model is shown in black.} 
\label{CPL fit to H(z)} 
\end{figure}

Figure \ref{CPL fit to H(z)} shows the reconstructed $H(z)$ (red curve) with $1$-$\sigma$ uncertainty band (light gray shaded area between dashed red curves), overlaid with the best-fit CPL model (black solid line). The CPL curve closely follows the reconstructed expansion for $z<1$, but approaches the edge of the $1$-$\sigma$ interval as z grow towards 1.5.

\subsection{Reconstructing $w(z)$ of Dark Energy}

\begin{figure}[h!]
\centering
\includegraphics[width=1 \columnwidth]{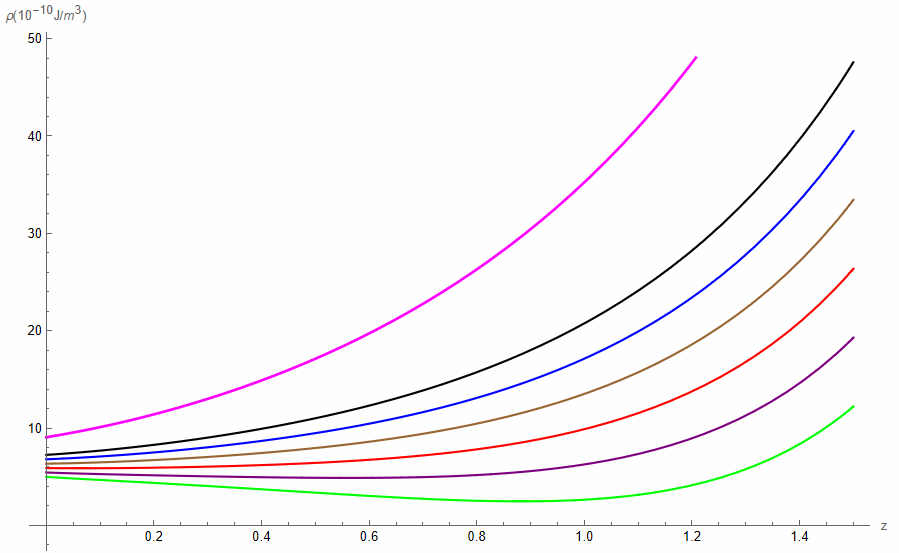}
\caption{Reconstructed average dark energy density $\rho_{\rm DE}(z)$ for different values of ${\rm \Omega}_m \in$ (0.2, 0.25, 0.3, 0.35, 0.4, 0.45). Each curve is obtained by subtracting the matter contribution $\rho(0) {\rm \Omega}_m (1+z)^3$ from the total average reconstructed $\rho(z)$. The curves are ordered from top to bottom according to increasing ${\rm \Omega}_m$.} 
\label{Average rho for different omega} 
\end{figure}

\begin{figure}[h!]
\centering
\includegraphics[width=1 \columnwidth]{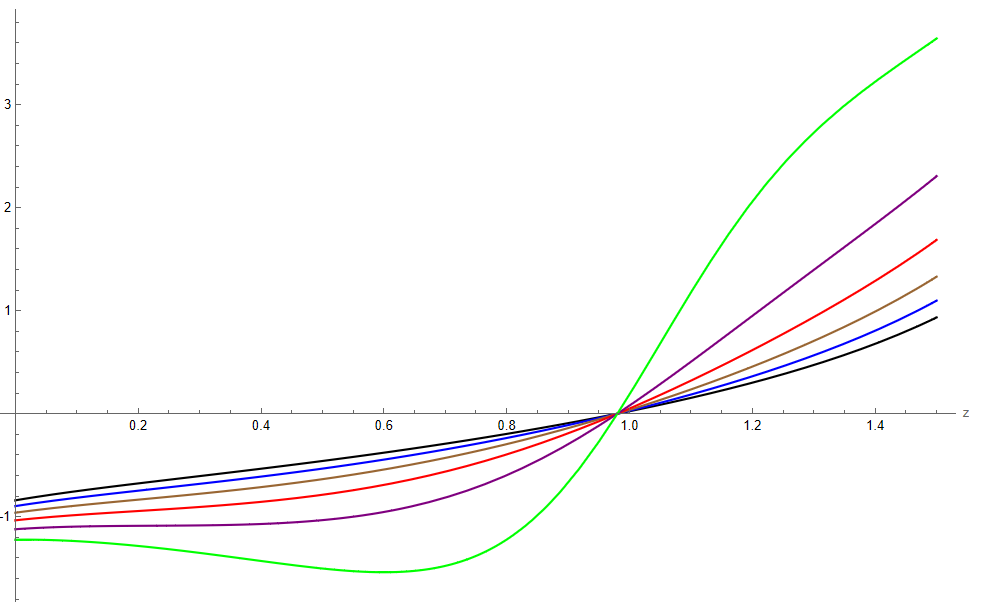}
\caption{Reconstructed equation of state parameter $w_{\rm DE}(z)$, obtained from the previously derived $\rho_{\rm DE}(z)$ and the reconstructed pressure $p(z)$. The curves correspond to the same ${\rm \Omega}_m$values as in Figure \ref{Average rho for different omega}: (0.2, 0.25, 0.3, 0.35, 0.4, 0.45), with colors matched accordingly.} 
\label{Average w for different omega} 
\end{figure}

\begin{figure}[h!]
\centering
\includegraphics[width=1 \columnwidth]{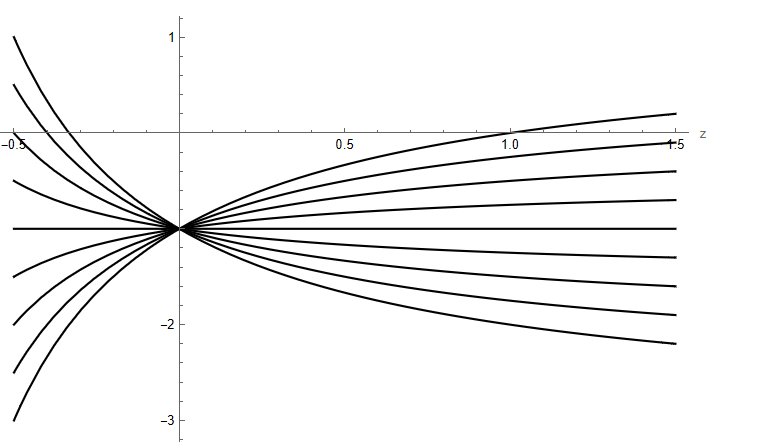}
\caption{CPL parametrization of the dark energy equation of state $w(z)$ for various values of $w_a$, with $w_0$ fixed at -1. The curves illustrate the impact of different $w_a$ values in the range {-2, -1.5, -1, -0.5, 0, 0.5, 1, 1.5, 2} on the redshift evolution of $w(z)$.} 
\label{CPL for different wa} 
\end{figure}

Our spread-LDF procedure gave us the total energy density of the universe as a function redshift via Eq. (\ref{rho}). We previously recalled that the matter energy density has the form (\ref{rho_matter_evolution}), and the balance is what is usually called dark energy, and used this relationship to derive an upper-bound on matter density and lower-bound on dark energy density in Section 4.1.  Now we want to focus on the mentioned balance, but we do not really know $\rho_m(0)$. The usually quoted ${\rm \Omega}_m$ value of $0.3$ comes from the concordance ${\rm \Lambda}$CDM model, and we do not want to make that assumption here. But we do have the upper-bound of 0.48 from Section 4.1 and a lower-bound of $\sim$0.2 from dynamical (non-cosmological) measurements up to the scale of galaxy clusters. So we decide to plot the density of dark energy as function of redshift for ${\rm \Omega}_m$ values 0.2, 0.25, 0.3, 0.35, 0.4 and 0.45  (Figure \ref{Average rho for different omega}) to see how we can proceed. We also plot the corresponding $w_{\rm DE}(z)$ curves  (Figure \ref{Average w for different omega}). The latter figure can be compared with the CPL $w(z)$ curves shown in Figure \ref{CPL for different wa}, where we held $w_0$ fixed (changing it would simply move all the corresponding graphs vertically up or down) and plotted curves for a range of $w_a$ values. It is apparent that none of the CPL $w(z)$ curves has similar qualitative features to any of the $w_{\rm DE}(z)$ curves. It is tempting to try an alternative $w_{\rm DE}(z)$ parametrization, (e.g. the curves of another parametrization that can be found in the literature, the Barboza-Alcaniz parametrization \cite{Barboza}, are also qualitatively dissimilar) but finding a reasonably simple one seems to be a prohibitive task. In fact, a $w_{\rm{DE}}(z)$ parametrization may not be the most “physical” approach, as we will discuss in the next section.

\subsection{Evidence for sign-change in the EoS of (Generalized) Dark Energy}

\begin{figure}[h!]
\centering
\includegraphics[width=1 \columnwidth]{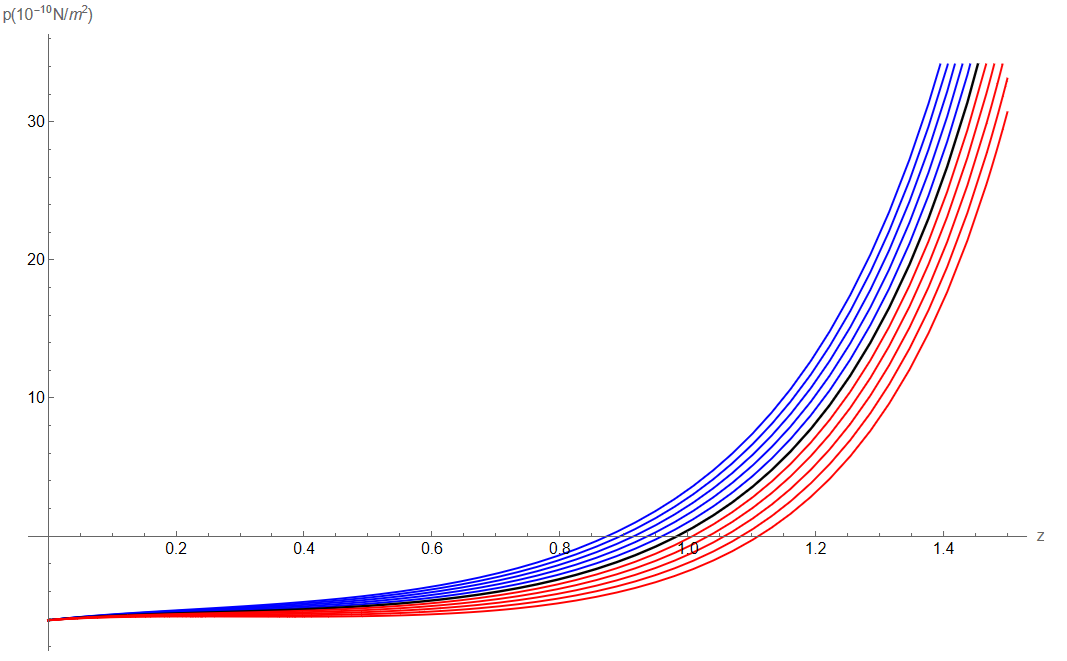}
\caption{Average reconstructed pressure for 11 values of the curvature parameter $\kappa$. Color encodes curvature as in Figures \ref{grid without GRBs}-\ref{average_acceleration_all_curves} \& \ref{Hubble average}-\ref{rho}.} 
\label{Pressure for all curves} 
\end{figure}

\begin{figure}[h!]
\centering
\includegraphics[width=1 \columnwidth]{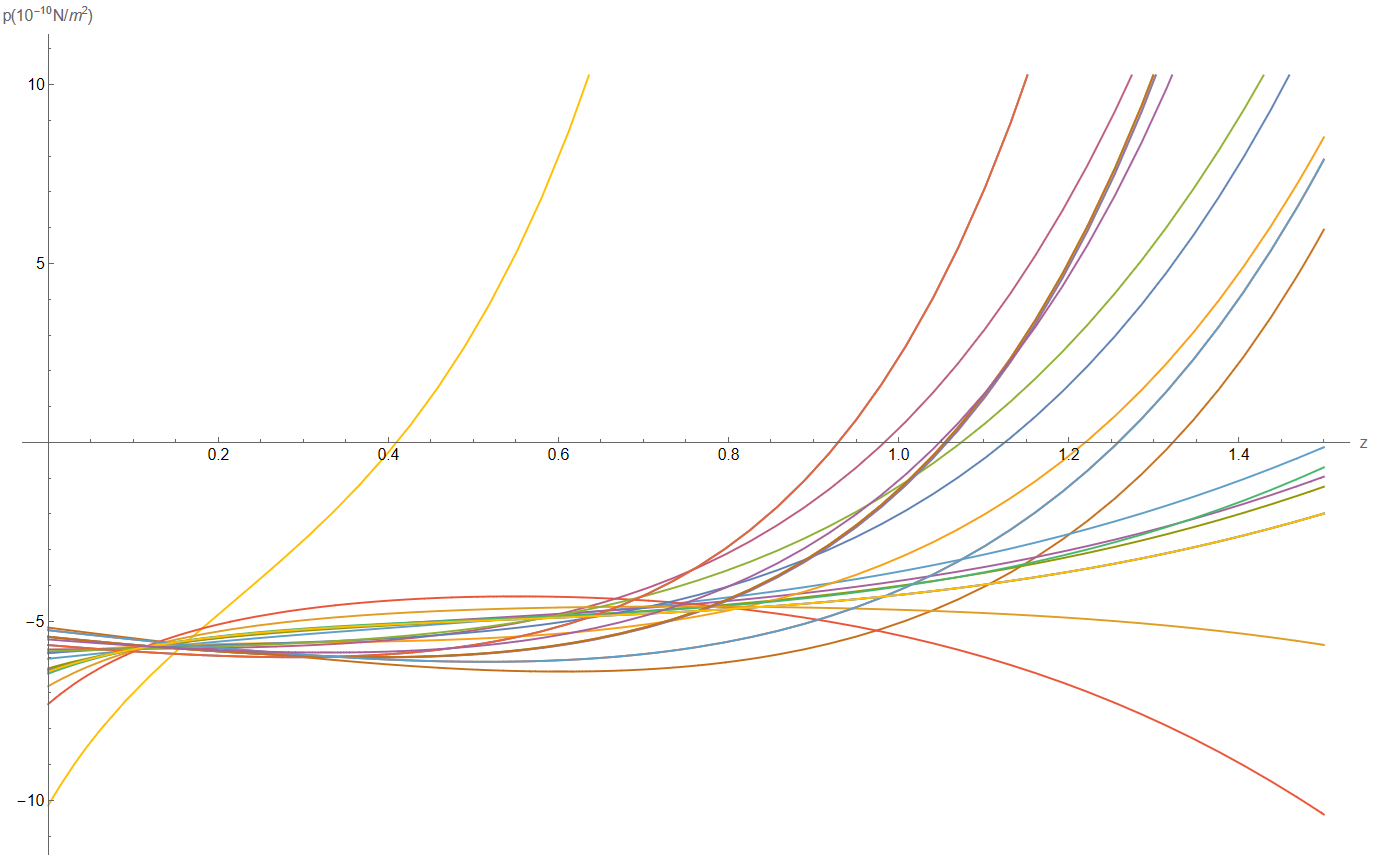}
\caption{Reconstructed pressure as a function of redshift for all selected functions for the spatially flat case.} 
\label{Pressure for all functions} 
\end{figure}

In producing the $w_{{\rm DE}}(z)$ curves  (Figure \ref{Average w for different omega}), we naturally used the expression of pressure, Eq. (\ref{frw pressure}), which we divided by the density of dark energy, similar to Eq. (\ref{simple EoS}). It turns out that the pressure (rather, the average of the pressures resulting from the spread of the fitted luminosity distance functions) changes from negative to positive at $\approx$0.98 (Figure \ref{Pressure for all curves}), hence, the sign of $w_{\rm DE}(z)$, will switch sign at the same redshift. This is the reason why all curves in Figure \ref{Average w for different omega} have the same $z$-intercept. Each curve has one intercept and no singularity, since we assume the energy density of dark energy to be always positive. In fact, this is how we derived an upper-bound on the present matter density; to get negative energy density of dark energy (around $z \approx 1$) we would have to assume an even larger ${\rm \Omega}_m$. In other words, it appears that adopting sign-changing dark energy \cite{Akarsu, Soriano} requires increasing ${\rm \Omega}_m$ substantially.
But positive pressure for $z>1$ seems to be a quite significant claim, not compatible with matter or cosmological constant. Radiation does have positive pressure, however radiation domination is supposed to have ended long ago, around $z\approx1000$. So let us look at the graphs more closely, and instead of looking at the average of the spread of 24 functions, let us plot them separately (Figure \ref{Pressure for all functions}). We can see that most curves do make a transition to positive pressure, although a few curves behave very differently from the rest (another illustration of the danger of getting trapped in a discordant model by making unlucky assumptions, without strong physical motivation). This increases our confidence in the transition from negative $p$ to positive $p$. However, it is desirable to continue this investigation for higher values of $z$, where SnIa data are not available. Recently, there have been significant efforts to standardize sources up to $z \approx 8$, GRB’s \cite{Daniotti} and quasars \cite{Elisabeta, Elisabeta2}, and we plan to do this in future work.

\subsubsection{Reconstructed Equation(s) of State for Generalized Dark Energy}

So far, we have discussed $w_{\rm DE}(z)$, however, the $w$ parameter is not epistemologically very meaningful if it is not constant. For example, in FLRW cosmology we can say that $\rho \approx a^{-3(1+w)}$  if $w$ is constant, but we cannot express $\rho$ in terms of $w$ directly if $w$ is variable. In fact, strictly speaking, the equation of state in physics is a relation between pressure and density (in most contexts, usually entropy per fluid particle is assumed constant). Until recently, in cosmology the ratio $p/\rho=w$ was taken constant and called the equation-of-state-parameter, but if it is not constant, as suspected in recent years, it is not very meaningful to use the ratio, one should default back to the relation. Expressing the relation in terms of $z$ does not seem to be very meaningful either: The relation $f(p,\rho)=0$ is a {\it local} relation, as befits a fluid consisting of particles, or by causality arguments; but z is not a local variable, there is no apparent reason that the local behaviour of a fluid should depend on $z$, the redshift, or indirectly, the age of the universe (time-translational invariance of laws of physics). One might argue that $z$ also determines the scale of the universe, or in non-flat cases, its curvature; the answer for the latter aspect is the equivalence principle, that local physics can be applied as if the universe is locally flat, and the answer for the former aspect is partially the same principle, partially the principle of causality/locality, that what happens far away in the universe cannot significantly affect what happens here.

\begin{figure}[h!]
\centering
\includegraphics[width=1 \columnwidth]{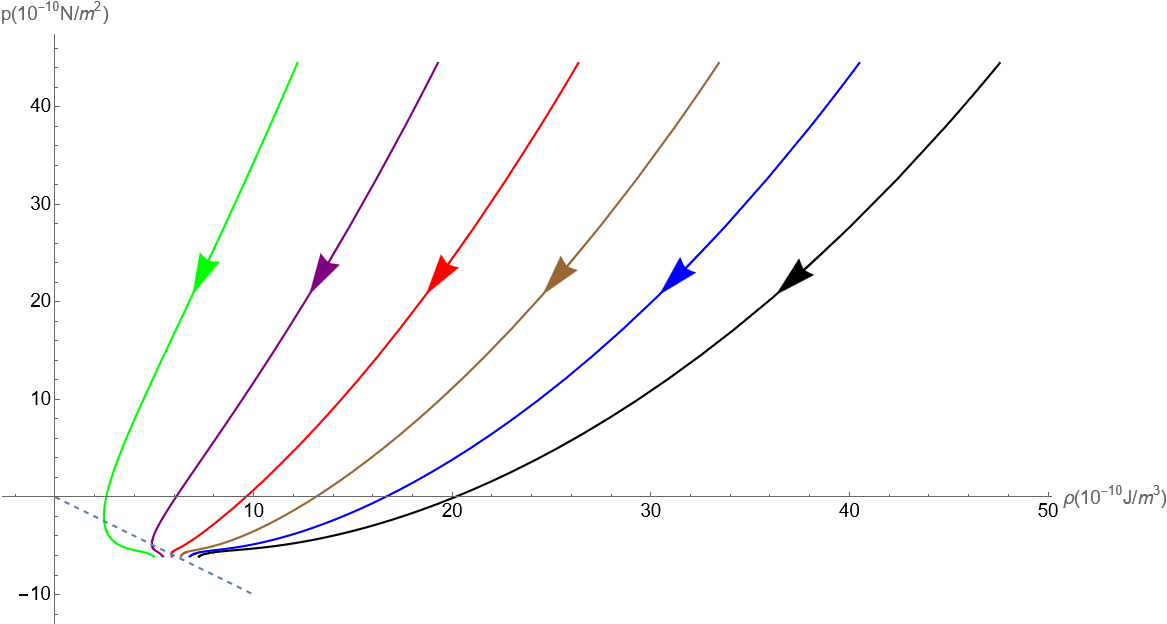}
\caption{Parametric function pressure in terms of dark energy density. The curves correspond to the same ${\rm \Omega}_m$ values as in Figure \ref{Average rho for different omega}: (0.2, 0.25, 0.3, 0.35, 0.4, 0.45), with colors matched accordingly. Arrows show decreasing $z$, i.e. time increasing towards the present; the dashed line corresponds to the phantom EoS $\rho+p=0$.} 
\label{p(rho)} 
\end{figure}

Hence we decide to investigate the relation between pressure and density of the non-matter part of the source of Einstein equations, as derived from the data. We decide to call that part “Generalized Dark Energy”, (GDE) since the concept of Dark Energy (DE) was invented to provide a source for accelerated expansion of the universe, hence has to satisfy $\rho+3 p < 0$, however, the source we find can have both energy density and pressure positive, so it can violate the DE property. So we plot in Figure \ref{p(rho)} parametrically $p$ as function of $\rho$, again for the different ${\rm \Omega}_m$ values used in Figure \ref{Average rho for different omega}. We attempted to find functions that would reproduce the dependence shown in the figure, at least for the single-valued curves, and quadratics gave reasonable fits (better than for example, an exponential version we tried), but the coefficients were inconsistent.

The arrows on the curves show decreasing $z$, i.e. show a flow from the past to the present. The dashed straight line represents the phantom EoS $\rho+p=0$, so the curves with large ${\rm \Omega}_m$ have already “crossed the phantom divide". The ${\rm \Omega}_m \approx 0.3$ curve, favored by the concordance model, has just arrived at the divide, the curves with smaller ${\rm \Omega}_m$ have not yet crossed it.

As discussed above, the disturbing possibility of a transition to positive pressure around $z\approx1$ should be rechecked with new and alternative data. If confirmed, one would have to look for models of sources that can account for the feature, maybe a kind of  dark radiation becoming dominant in that redshift range. But coming up with different ad hoc sources for different epochs would look epicyclical, so either a well-motivated particle/field model should be found to explain the GDE EoS over the whole range, or the assumptions leading to this result should be questioned: The Cosmological Principle, leading to LTB-like models or backreaction arguments, that is, effects of structure formation on assumed large-scale metric; the standardization of the candles \cite{Chung}, since they are the cornerstone observational support of the accelerating expansion concept; or the theory of gravity, namely, GR, leading to modified-gravity theories.

\section{Conclusion}

In this work, we have updated and re-applied a model-independent framework —which we refer to now as the spread-luminosity distance fitting (spread-LDF) method— for reconstructing the expansion history of the universe from observational data with minimal theoretical assumptions; in other words, the approach of going back from data to expansion history or information about contents, of the universe. The process, featuring a diverse family of functional ansätze across several redshift reparameterizations and selection of optimal function models via the Bayesian Information Criterion (BIC), was applied to the more recent, and significantly improved Pantheon+ Type Ia supernova dataset alongside a rescaled gamma-ray burst (GRB) sample to  enhance the stability of reconstructions. The BIC optimization in this incarnation was useful to avoid overfitting.

Our updated reconstruction confirms a robust transition from decelerated to accelerated cosmic expansion at a redshift of approximately $z_T \approx 0.44^{+0.04}_{-0.03}$, assuming flat spatial geometry. This key result is consistent with the acceleration inferred in standard ${\rm \Lambda}$CDM cosmology, but it emerges here entirely from the data and geometric structure of the FLRW metric, without invoking a specific matter or energy content. By analytically deriving the Hubble parameter and acceleration function from the reconstructed luminosity distance-redshift relation, we obtained a direct picture of the universe’s expansion history.

We then introduced a semi-model-independent approach by assuming the validity of General Relativity, while still refraining from adopting a detailed model of the universe's matter-energy content. Through this, we derived effective energy density $\rho(z)$, pressure $p(z)$, and equation of state parameter $w(z)$ from the reconstructed expansion. Our results indicate that the observed acceleration cannot be explained by matter alone. Specifically, we found that the energy density significantly deviates from the matter-dominated scaling $\rho_m \propto (1+z)^3$ at low redshifts, requiring a component with negative pressure—consistent with dark energy or modified gravity effects. Quantitatively, we derived bounds such as ${\rm \Omega}_m^{max}\approx 0.48$ and ${\rm \Omega}_{\rm \Lambda}^{min}\approx 0.52$ , in good agreement with ${\rm \Lambda}$CDM expectations, but obtained without assuming a cosmological constant. We note that to have dark energy with negative density at any point in our redshift range, as proposed in some works, ${\rm \Omega}_m$ needs to be $\sim 0.55$ or larger, i.e. exacerbates the dark matter problem.

Moreover, we also reconstructed the effective equation-of-state parameter $w(z)$. Our findings show that while the CPL parametrization agrees with the reconstructed curve within the estimated errors, it fails to reproduce the qualitative behaviour of the reconstructed $w(z)$, especially since we find that the pressure becomes positive for $z>~{\sim1}$. This mismatch emphasizes the limitations of commonly used dark energy parameterizations when interpreted too literally and highlights the utility of our concept of going back from the observations to contents of the universe, rather than from assumed contents to the observations. 

We can also make plots of $p(\rho)$ for different values of ${\rm \Omega}_m$. The plots do not seem to suggest a simple analytic formula for the EoS, but indicate a propensity to approach/cross the phantom divide, especially for larger values of ${\rm \Omega}_m$. More significantly, the finding of positive pressure for the cosmic fluid at $z>~{\sim1}$, if confirmed by future studies, will have profound implications: It will lead to either questioning of the application of the Cosmological Principle to our universe; or to stronger scrutiny of observational/mathematical effects of cosmic structure formation; or development of more complex models for Generalized Dark Energy; or alternatively, stronger motivation for modified/alternative gravity theories.

\section{Acknowledgements}  

We are grateful to V. Erkcan \" Ozcan for helpful discussions.

\end{document}